\documentclass[letterpaper,12pt]{article}
\usepackage[margin=1in]{geometry}
\usepackage{verbatim}
\usepackage{amsmath}
\usepackage{amssymb}
\usepackage{graphicx}
\usepackage{amsthm}
\usepackage{subfig}
\usepackage[round,authoryear]{natbib}

\def\epsilon{\varepsilon}
\def\bigoh{\hat{P}}
\def\oh{\hat{p}}

\def\phi{\varphi}

\def\0s{{\bf 0}}

\newtheorem{theorem}{Theorem}[section]

\newtheorem{observe}[theorem]{Observation}
\newtheorem{remark1}[theorem]{Remark}

\newenvironment{remark}{\begin{remark1} \rm}{\end{remark1}}

\title{\Large{An introduction to how $\chi^2$ and classical exact tests
       often wildly misreport significance
       and how the remedy lies in computers}\thanks{An extended version
       of the present article is available at http://arxiv.org/abs/1108.4126}}
\author{William Perkins\thanks{wperkins3@math.gatech.edu,
        Georgia Institute of Technology, 686 Cherry St.,
        Atlanta, GA 30332-0160},
        Mark Tygert\thanks{tygert@courant.nyu.edu,
        Courant Institute, NYU, 251 Mercer St., New York, NY 10012},
        and Rachel Ward\thanks{rward@math.utexas.edu,
        University of Texas, 1 University Station, C1200, Austin, TX 78712}}

\begin{document}

\maketitle

\begin{abstract}
\noindent
Goodness-of-fit tests based on the Euclidean distance
often outperform $\chi^2$ and other classical tests
(including the standard exact tests) by at least an order of magnitude
when the model being tested for goodness-of-fit
is a discrete probability distribution that is not close to uniform.
The present article discusses numerous examples of this.
Goodness-of-fit tests based on the Euclidean metric are now practical
and convenient: although the actual values taken by the Euclidean distance
and similar goodness-of-fit statistics are seldom humanly interpretable,
black-box computer programs can rapidly calculate their precise significance.

\medskip
\smallskip

\end{abstract}

\nocite{hollander-wolfe}
\nocite{mayo}
\nocite{wasserman}

\tableofcontents

\section{Introduction}
\label{intro}

A basic task in statistics is to ascertain whether a given set
of independent and identically distributed (i.i.d.)\ draws does not come
from a given ``model,'' where the model may consist of either
a single fully specified probability distribution or
a parameterized family of probability distributions.
The present paper concerns the case in which the draws
are discrete random variables, taking values in a finite or countable set.
In accordance with the standard terminology,
we will refer to the possible values of the discrete random variables
as ``bins'' (``categories,'' ``cells,'' and ``classes'' are common synonyms
for ``bins'').

A natural approach to ascertaining whether the i.i.d.\ draws
do not come from the model uses a root-mean-square statistic.
To construct this statistic,
we estimate the probability distribution over the bins
using the given i.i.d.\ draws, and then measure	the root-mean-square
difference between this empirical distribution and the model distribution
\citep[see, for example,][page 123; or Section~\ref{definitions} below]
      {rao,varadhan-levandowsky-rubin}.
If the draws do in fact arise from the model,
then with high probability this root-mean-square is not large.
Thus, if the root-mean-square statistic is large,
then we can be confident that the draws do not arise from the model.

To quantify ``large'' and ``confident,''
let us denote by $x$ the value of the root-mean-square
for the given i.i.d.\ draws;
let us denote by $X$ the root-mean-square statistic
constructed for different i.i.d.\ draws that definitely do in fact come
from the model (if the model is parameterized, then we draw
from the distribution corresponding to the parameter given
by a maximum-likelihood estimate for the experimental data).
The ``P-value'' $P$ is then defined to be the probability
that $X \ge x$ (viewing $X$ --- but not $x$ --- as a random variable).
Given the P-value $P$, we can have $100(1-P)$\% confidence that the draws
do not arise from the model.

Now, the P-values for the simple root-mean-square statistic can be
different functions of $x$ for different model probability distributions.
To avoid this seeming inconvenience asymptotically
(in the limit of large numbers of draws),
K. Pearson replaced the uniformly weighted mean in the root-mean-square
with a weighted average; the weights are the reciprocals
of the model probabilities associated with the various bins.
This produces the classic~$\chi^2$ statistic introduced by \cite{pearson}
--- see, for example, formula~(\ref{chi2}) below.
However, when model probabilities can be small
(relative to others in the same model distribution),
this weighted average can involve division by nearly zero.
As demonstrated below, dividing by nearly zero severely restricts
the statistical power of $\chi^2$
--- even in the absence of round-off errors ---
especially when dividing by nearly zero for each of many bins.
The problem arises whether or not every bin contains several draws
(see Remark~\ref{asymps}). \cite{press} tackled similar issues.

The main thesis of the present article is that using only the classic $\chi^2$
statistic is no longer appropriate, that certain alternatives are far superior
now that computers are widely available.
As illustrated below, the simple root-mean-square, used in conjunction
with the log--likelihood-ratio ``$G^2$'' goodness-of-fit statistic,
is generally preferable to the classic $\chi^2$ statistic.
(The log--likelihood-ratio also involves division by nearly zero,
but tempers this somewhat by taking a logarithm.)
We do not make any claim that this is the best possible alternative.
In fact, the discrete Kolmogorov-Smirnov and related statistics
used by~\cite{clauset-shalizi-newman} and~\cite{dagostino-stephens}
can be more powerful than the root-mean-square in certain circumstances;
in any case, the discrete Kolmogorov-Smirnov statistic and the root-mean-square
are similar in many ways, and complementary in others.
We focus on the root-mean-square largely because it is so simple
and easy to understand; for example, computing the P-values
of the root-mean-square in the limit of large numbers of draws is trivial,
even when estimating continuous parameters via maximum-likelihood methods,
as discussed by \cite{perkins-tygert-ward1,perkins-tygert-ward2}.
Furthermore, the classic $\chi^2$ statistic is just a weighted version
of the root-mean-square, facilitating their comparison.
Finally, $\chi^2$ and the root-mean-square coincide
when the model distribution is uniform.

Please note that all statistical tests reported in the present paper
(including those involving the $\chi^2$ statistic) are exact;
we compute P-values via Monte-Carlo simulations
providing guaranteed error bounds (see Section~\ref{hypotheses} below).
In all numerical results reported below,
we generated random numbers via the C programming language procedure given
on page~9 of~\cite{marsaglia},
implementing the recommended complementary multiply with carry.

To be sure, the problem with $\chi^2$ is neither subtle nor esoteric.
For a particularly revealing example, see Subsection~\ref{hardyweinberg} below.

Appropriate rebinning to uniformize the probabilities associated with the bins
can mitigate much of the problem with $\chi^2$.
Yet rebinning is a black art that is liable to improperly influence the result
of a goodness-of-fit test. Moreover, rebinning requires careful extra work,
making $\chi^2$ less easy-to-use. A principal advantage of the root-mean-square
is that it does not require any rebinning; indeed, the root-mean-square
is most powerful without any rebinning.

\begin{remark}
\label{asymps}
In many of our examples, there is a bin
for which the expected number of draws is very small under the model.
Please note that, although it is natural for the expected numbers of draws
for some bins to be very small, especially when the model has many bins,
the advantage of the root-mean-square over $\chi^2$ is substantial
even when the expected number of draws is at least five for every bin;
see, for example, Subsection~\ref{firstex} or Subsection~\ref{sixthexp}.
\end{remark}

\begin{remark}
\label{goodness}
Goodness-of-fit tests are probably most useful in practice
not for ascertaining whether a model is correct or not,
but for determining whether the discrepancy between the model
and experiment is larger than expected random fluctuations.
While models outside the physical sciences typically are not exactly correct,
testing the validity of using a model for virtually any purpose requires
knowing whether observed discrepancies are due to inaccuracies
or inadequacies in the models or (on the contrary) could be due
to chance arising from necessarily finite sample sizes.
Thus, goodness-of-fit tests
are critical even when the models are not supposed to be exactly correct,
in order to gauge the size of the unavoidable random fluctuations.
For further clarification, see the remarkably extensive title used
by~\cite{pearson} introducing $\chi^2$;
see also the modern treatments by~\cite{gelman} and~\cite{cox}.
\end{remark}

\begin{remark}
Combining the root-mean-square methodology and the statistical bootstrap
given by~\cite{efron-tibshirani} should produce a test
for whether two separate sets of draws arise from the same
or from different distributions, when each set is taken i.i.d.\
from some (unspecified) distribution; the two distributions associated
with the sets may differ.
This is related to testing for association/independence/homogeneity
in contingency-tables/cross-tabulations that have only two rows.
\end{remark}

\section{Definitions of the test statistics}
\label{definitions}

In this section, we review the definitions of four goodness-of-fit statistics
--- the root-mean-square, $\chi^2$, the log--likelihood-ratio or $G^2$,
and the Freeman-Tukey or Hellinger distance.
The latter three statistics are the best-known members
of the standard Cressie-Read power-divergence family,
as discussed by~\cite{rao}.
We use $p_0^{(1)}$,~$p_0^{(2)}$, \dots,~$p_0^{(m)}$
to denote the modeled fractions
of $n$ i.i.d.\ draws falling in $m$ bins, numbered $1$,~$2$, \dots,~$m$,
respectively, and we use $\oh^{(1)}$,~$\oh^{(2)}$, \dots,~$\oh^{(m)}$
to denote the observed fractions
of the $n$ draws falling in the respective bins.
That is, $p_0^{(1)}$,~$p_0^{(2)}$, \dots,~$p_0^{(m)}$ are the probabilities
associated with the respective bins in the {\it model} distribution,
whereas $\oh^{(1)}$,~$\oh^{(2)}$, \dots,~$\oh^{(m)}$ are the fractions
of the $n$ draws falling in the respective bins when we take the draws
from a distribution that may differ from the model
--- their {\it actual} distribution.
Specifically, if $i_1$,~$i_2$, \dots,~$i_n$ are
the observed i.i.d.\ draws, then $\oh^{(j)}$ is $\frac{1}{n}$ times
the number of $i_1$,~$i_2$, \dots,~$i_n$ falling in bin~$j$,
for $j = 1$,~$2$, \dots,~$m$.
If the model is parameterized by a parameter $\theta$,
then the probabilities $p_0^{(1)}$,~$p_0^{(2)}$, \dots,~$p_0^{(m)}$
are functions of $\theta$;
if the model is fully specified, then we can view
the probabilities $p_0^{(1)}$,~$p_0^{(2)}$, \dots,~$p_0^{(m)}$ as constant
as functions of~$\theta$.
We use $\hat\theta$ to denote a maximum-likelihood estimate of~$\theta$
obtained from $\oh^{(1)}$,~$\oh^{(2)}$, \dots,~$\oh^{(m)}$.

With this notation, the root-mean-square statistic is
\begin{equation}
\label{rms}
x = \sqrt{\frac{1}{m} \sum_{j=1}^m (\oh^{(j)} - p_0^{(j)}(\hat\theta))^2}.
\end{equation}
We use the designation ``root-mean-square'' to refer to $x$.

The classical Pearson $\chi^2$ statistic is
\begin{equation}
\label{chi2}
\chi^2 = n \sum_{j=1}^m \frac{(\oh^{(j)} - p_0^{(j)}(\hat\theta))^2}
                             {p_0^{(j)}(\hat\theta)},
\end{equation}
under the convention that
$(\oh^{(j)} - p_0^{(j)}(\hat\theta))^2/p_0^{(j)}(\hat\theta) = 0$
if $p_0^{(j)}(\hat\theta) = 0 = \oh^{(j)}$.
We use the standard designation ``$\chi^2$'' to refer to $\chi^2$.

The log--likelihood-ratio or ``$G^2$'' statistic is
\begin{equation}
\label{lr}
g^2 = 2n \sum_{j=1}^m \oh^{(j)}
\, \ln\left( \frac{\oh^{(j)}}{p_0^{(j)}(\hat\theta)} \right),
\end{equation}
under the convention that
$\oh^{(j)} \, \ln(\oh^{(j)}/p_0^{(j)}(\hat\theta)) = 0$
if $\oh^{(j)} = 0$.
We use the common designation ``$G^2$'' to refer to $g^2$.

The Freeman-Tukey or Hellinger-distance statistic is
\begin{equation}
\label{freeman-tukey}
h^2
= 4n \sum_{j=1}^m \left(\sqrt{\oh^{(j)}} - \sqrt{p_0^{(j)}(\hat\theta)}\right)^2
= 4n \sum_{j=1}^m \left[ (\oh^{(j)} - p_0^{(j)}(\hat\theta))^2 \bigg/
  \left(\sqrt{\oh^{(j)}} + \sqrt{p_0^{(j)}(\hat\theta)}\right)^2 \right].
\end{equation}
We use the well-known designation ``Freeman-Tukey'' to refer to $h^2$.

In the limit that the number $n$ of draws is large,
the distributions of $\chi^2$~defined in~(\ref{chi2}),
$g^2$~defined in~(\ref{lr}), and $h^2$~defined in~(\ref{freeman-tukey})
are all the same when the actual underlying distribution of the draws
comes from the model, as discussed, for example, by~\cite{rao}.
However, when the number $n$ of draws is not large,
then their distributions can differ substantially.
In all our data and power analyses, we compute P-values
via Monte-Carlo simulations, without relying on the number $n$ of draws
to be large.

\section{Hypothesis tests with parameter estimation}
\label{hypotheses}

In this section, we discuss the testing of hypotheses
involving parameterized models:
Given a family $p_0(\theta)$ of probability distributions
parameterized by $\theta$, and given observed i.i.d.\ draws
from some actual underlying (unknown) distribution $p$,
we would like to test the hypothesis
\begin{equation}
\label{null'}
H'_0 : \hbox{for some } \theta,\; p = p_0(\theta),
\end{equation}
against the alternative
\begin{equation}
H'_1 : \hbox{for all } \theta,\; p \ne p_0(\theta).
\end{equation}
Given only finitely many draws,
the P-value for such a test would have to be independent
of the parameter $\theta$, since the proper value for $\theta$ is unknown
($\theta$ is known as a ``nuisance'' parameter).
Unfortunately, it is not clear how to devise such a test
when the probability distributions are discrete.
None of the standard methods (including $\chi^2$,
the log--likelihood-ratio, the Freeman-Tukey/Hellinger distance,
and other Cressie-Read power-divergence statistics) produce P-values
that are independent of the parameter $\theta$.
Some methods do produce P-values that are independent of $\theta$
in the limit of large numbers of draws, but this is not especially useful,
since in the limit of large numbers of draws any actual parameter $\theta$
would be almost surely known exactly anyway;
further elaboration is available in Appendix~B of~\cite{perkins-tygert-ward3}.

In the present paper, we test the significance of assuming
\begin{equation}
\label{null}
H_0 : p = p_0(\hat\theta) \hbox{ for the particular observed value
of $\hat\theta$},
\end{equation}
where $\hat\theta$ is a maximum-likelihood estimate of $\theta$;
i.e., $H_0$ is the hypothesis that $p = p_0(\hat\theta)$
for the value of $\hat\theta$ associated
with the single realization of the experiment that was measured
(subsequent repetitions of the experiment, including those considered
when calculating the P-value as in Remark~\ref{MonteCarlo},
can yield different estimates of the parameter,
even though the repetitions' actual distribution $p$ is the same).
Of course, the accuracy of the estimate~$\hat\theta$~generally improves
as the number of draws increases;
testing~(\ref{null'}) and testing~(\ref{null}) are asymptotically equivalent,
in the limit of large numbers of draws, under the conditions of~\cite{romano}.

As testing exactly $H'_0$ defined in~(\ref{null'}) does not seem
to be feasible in general when the probability distributions are discrete
and there are more than just a few bins,
we focus on testing the closely related assumption $H_0$ defined
in~(\ref{null}). The latter is more relevant for many applications, anyways
--- plots typically display the particular fitted distribution in~(\ref{null});
interpreting such plots naturally involves~(\ref{null}).
All tests of the present paper concern the significance
of assuming~$H_0$ defined in~(\ref{null}) (if the model is fully specified,
then the probability distribution $p_0(\theta)$ is the same for all $\theta$).
Please be sure to bear in mind Remark~\ref{goodness} of Section~\ref{intro}.
{\it A significance test simply gauges the consistency of the observed data
with our assumption}; we are not trying to decide whether the assumption is
likely to be true~(or~false), nor are we trying to decide whether some
alternative assumption is likely to be true~(or~false).

\begin{remark}
Another means of handling nuisance parameters is to test the hypothesis
\begin{equation}
\label{null''}
H''_0 : p = p_0(\hat\theta) \hbox{ for all possible realizations
of the experiment};
\end{equation}
that is, $H''_0$ is the hypothesis that $p = p_0(\hat\theta)$
and that $p_0(\hat\theta)$ always takes exactly the same value
during repetitions of the experiment.
The assumption that~(\ref{null''}) is true seems to be more extreme,
a more substantial departure from~(\ref{null'}), than~(\ref{null}).
Nevertheless, testing~(\ref{null''}) is standard;
see, for example, Section~6 of~\cite{cochran}.
Assuming~(\ref{null''}) amounts to conditioning~(\ref{null'})
on a statistic that is minimally sufficient for estimating $\theta$;
computing the associated P-values is not always trivial.
Testing the significance of assuming~(\ref{null}) would seem
to be more apropos in practice for applications
in which the experimental design does not enforce
that repeated experiments always yield the same value for $p_0(\hat\theta)$.
\end{remark}

\begin{remark}
The parameter $\theta$ can be integer-valued, real-valued, complex-valued,
vector-valued, matrix-valued, or any combination of the many possibilities.
For instance, when we do not know the proper ordering of the bins a priori,
we must include a parameter that contains a permutation
(or permutation matrix) specifying the order of the bins;
maximum-likelihood estimation then entails
sorting the model and all empirical frequencies
(whether experimental or simulated) ---
see Subsection~\ref{zipfsec} for details.
With Remark~\ref{MonteCarlo}, we need not contemplate
how many degrees of freedom are in a permutation.
\end{remark}

\begin{remark}
\label{MonteCarlo}
To compute the P-value assessing the consistency
of the experimental data with assuming~(\ref{null}),
we can use Monte-Carlo simulations
(very similar to those used by~\cite{clauset-shalizi-newman}).
First, we estimate the parameter $\theta$
from the $n$ given experimental draws, obtaining $\hat\theta$,
and calculate the statistic
($\chi^2$, $G^2$, Freeman-Tukey, or the root-mean-square),
using the given data and taking the model distribution to be $p_0(\hat\theta)$.
We then run many simulations.
To conduct a single simulation, we perform the following three-step procedure:
\begin{enumerate}
\item we generate $n$ i.i.d.\ draws according
      to the model distribution $p_0(\hat\theta)$,
      where $\hat\theta$ is the estimate calculated
      from the experimental data,
\item we estimate the parameter $\theta$ from the data
      generated in Step~1, obtaining a new estimate $\tilde\theta$, and
\item we calculate the statistic under consideration
      ($\chi^2$, $G^2$, Freeman-Tukey, or the root-mean-square),
      using the data generated in Step~1 and
      taking the model distribution to be $p_0(\tilde\theta)$,
      where $\tilde\theta$ is the estimate calculated in Step~2
      from the data generated in Step~1.
\end{enumerate}
After conducting many such simulations, we may estimate the P-value
for assuming~(\ref{null})
as the fraction of the statistics calculated in Step~3
that are greater than or equal to the statistic calculated
from the empirical data.
The accuracy of the estimated P-value
is inversely proportional to the square root
of the number of simulations conducted;
for details, see Remark~\ref{error-bars} below.
This procedure works since, by definition, the P-value is the probability that
\begin{equation}
\label{theevent}
d\left[\left(\begin{array}{c}\bigoh^{(1)}\\\bigoh^{(2)}\\\vdots\\
\bigoh^{(m)}\end{array}\right),
\left(\begin{array}{c}p_0^{(1)}(\hat\Theta)\\p_0^{(2)}(\hat\Theta)\\\vdots\\
p_0^{(m)}(\hat\Theta)\end{array}\right)\right] \quad\ge\quad
d\left[\left(\begin{array}{c}\oh^{(1)}\\\oh^{(2)}\\\vdots\\
\oh^{(m)}\end{array}\right),
\left(\begin{array}{c}p_0^{(1)}(\hat\theta)\\p_0^{(2)}(\hat\theta)\\\vdots\\
p_0^{(m)}(\hat\theta)\end{array}\right)\right],
\end{equation}
where
\begin{itemize}
\item $m$ is the number of all possible values that the draws can take,
\item $d$ is the measure of the discrepancy
between two probability distributions
over $m$ bins (i.e., between two vectors each with $m$ entries)
that is associated with the statistic under consideration
($d$ is the Euclidean distance for the root-mean-square,
a weighted Euclidean distance for $\chi^2$,
the Hellinger distance for the Freeman-Tukey statistic,
and the relative entropy --- the Kullback-Leibler divergence ---
for the log--likelihood-ratio),
\item $\oh^{(1)}$,~$\oh^{(2)}$, \dots,~$\oh^{(m)}$ are the fractions
of the $n$ given experimental draws falling in the respective bins,
\item $\hat\theta$ is the estimate of $\theta$
obtained from $\oh^{(1)}$,~$\oh^{(2)}$, \dots,~$\oh^{(m)}$,
\item $\bigoh^{(1)}$,~$\bigoh^{(2)}$, \dots,~$\bigoh^{(m)}$
are the fractions of $n$ i.i.d.\ draws falling in the respective bins
when taking the draws from the distribution $p_0(\hat\theta)$
assumed in~(\ref{null}), and
\item $\hat\Theta$ is the estimate of the parameter~$\theta$
obtained from $\bigoh^{(1)}$,~$\bigoh^{(2)}$, \dots,~$\bigoh^{(m)}$
(note that $\hat\Theta$ is not necessarily always equal to $\hat\theta$:
even under the null hypothesis, repetitions of the experiment
could yield different estimates of the parameter;
see also Remark~\ref{tricky}).
\end{itemize}
When taking the probability that (\ref{theevent}) occurs,
only the left-hand side is random ---
we regard the left-hand side of~(\ref{theevent}) as a random variable
and the right-hand side as a fixed number
determined via the experimental data.
As with any probability, to compute the probability
that (\ref{theevent}) occurs,
we can calculate many independent realizations of the random variable
and observe that the fraction which satisfy~(\ref{theevent})
is a good approximation to the probability
when the number of realizations is large;
Remark~\ref{error-bars} details the accuracy of the approximation.
(The procedure in the present remark follows this prescription
to estimate P-values.)
\end{remark}

\begin{remark}
\label{error-bars}
The standard error of the estimate from Remark~\ref{MonteCarlo}
for an exact P-value $P$
is $\sqrt{P(1-P)/\ell}$, where $\ell$ is the number
of Monte-Carlo simulations conducted to produce the estimate.
Indeed, each simulation has probability $P$ of producing a statistic
that is greater than or equal to the statistic corresponding
to an exact P-value of $P$.
Since the simulations are all independent, the number of the $\ell$ simulations
that produce statistics greater than or equal to that corresponding
to P-value $P$ follows the binomial distribution
with $\ell$ trials and probability $P$ of success in each trial.
The standard deviation of the number of simulations whose statistics
are greater than or equal to that corresponding to P-value $P$ is therefore
$\sqrt{\ell P (1-P)}$, and so the standard deviation
of the {\it fraction} of the simulations producing such statistics
is $\sqrt{P (1-P)/\ell}$. Of course, the fraction itself
is the Monte-Carlo estimate of the exact P-value
(we use this estimate in place of the unknown $P$
when calculating the standard error $\sqrt{P (1-P)/\ell}$).
\end{remark}

\begin{remark}
\label{tricky}
For any family $p_0(\theta)$ of discrete probability distributions
parameterized by a permutation $\theta$
that specifies the order of the bins
(meaning that there exists a discrete probability distribution $q$ such that
$p_0^{(j)}(\theta) = q^{(\theta(j))}$ for all $j$),
and for any number $n$ of draws,
the P-values defined in Remark~\ref{MonteCarlo}
have the following highly desirable property:
Suppose that the actual underlying distribution $p$
of the experimental draws is equal to $p_0(\theta)$
for some (unknown) $\theta$.
Suppose further that $P$ is the P-value for assuming~(\ref{null}),
calculated for a particular realization of the experiment.
Consider repeating the same experiment over and over,
and calculating the P-value for each realization,
each time using that realization's particular maximum-likelihood estimate
of the parameter in the hypothesis~(\ref{null}).
Then, the fraction of the P-values that are greater than or equal to $P$
is equal to $P$ in the limit of many repetitions of the experiment.
This property is a compelling reason to use $d(\hat{P},p_0(\hat\Theta))$
rather than $d(\hat{P},p_0(\hat\theta))$
in the left-hand side of~(\ref{theevent}).
Also, the procedure of Remark~\ref{MonteCarlo} can be viewed as
a parametric bootstrap approximation,
as discussed, for example, by~\cite{romano}, \cite{efron-tibshirani},
and~\cite{bickel-ritov-stoker}:

For any family $p_0(\theta)$ of discrete probability distributions,
the P-values defined in Remark~\ref{MonteCarlo}
have the following additional highly desirable property:
Suppose that the actual underlying distribution $p$
of the experimental draws is equal to $p_0(\theta)$ for some (unknown) $\theta$.
Consider repeating the experiment over and over,
and calculating the P-value for each realization,
each time using that realization's particular maximum-likelihood estimate
of the parameter in the hypothesis~(\ref{null}).
Then, the resulting P-values converge in distribution
to the uniform distribution over $(0,1)$,
in the limit of large numbers of draws.

It may be somewhat fortuitous that the scheme in Remark~\ref{MonteCarlo}
has so many favorable properties; indeed,
\cite{bayarri-berger} and~\cite{robins-vanderwaart-ventura} (among others)
have pointed out problems with certain generalizations.
\end{remark}

\section{Data analysis}

In this section, we use several data sets to investigate the performance
of goodness-of-fit statistics.
The root-mean-square generally performs much better
than the classical statistics.
We take the position that a user of statistics should not have to worry
about rebinning; we discuss rebinning only briefly.
We compute all P-values via Monte Carlo as in Remark~\ref{MonteCarlo};
Remark~\ref{error-bars} details
the guaranteed accuracy of the computed P-values.

\subsection{Synthetic examples}

To better explicate the performance of the goodness-of-fit statistics,
we first analyze some toy examples.
We consider the model distribution
\begin{equation}
\label{synth1}
p_0^{(1)} = \frac{1}{4},
\end{equation}
\begin{equation}
\label{synth2}
p_0^{(2)} = \frac{1}{4},
\end{equation}
and
\begin{equation}
\label{synth3}
p_0^{(j)} = \frac{1}{2m-4}
\end{equation}
for $j = 3$,~$4$, \dots,~$m$.
For the empirical distribution, we first use $n = 20$ draws,
with 15 in the first bin, 5 in the second bin, and no draw in any other bin.
This data is clearly unlikely to arise from the model
specified in~(\ref{synth1})--(\ref{synth3}), but we would like to see exactly
how well the various goodness-of-fit statistics detect the obvious discrepancy.

\begin{figure}
\begin{center}
\rotatebox{-90}{\scalebox{.47}{\includegraphics{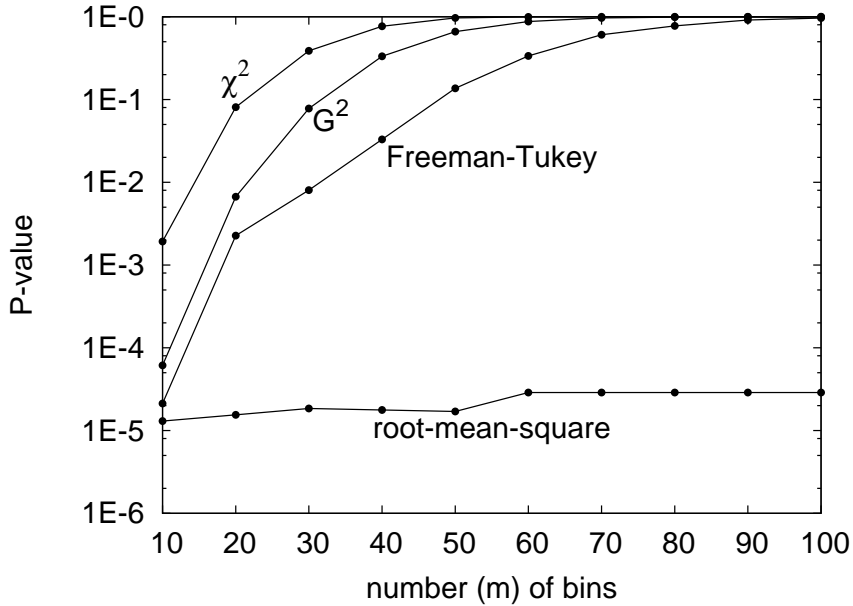}}}
\\\vspace{.1in}
\caption{P-values for the hypothesis
         that the model~(\ref{synth1})--(\ref{synth3}) agrees
         with the data of 15 draws in the first bin,
         5 draws in the second bin, and no draw in any other bin}
\label{plotsynth}
\end{center}
\end{figure}

Figure~\ref{plotsynth} plots the P-values for testing
whether the empirical data arises from the model
specified in~(\ref{synth1})--(\ref{synth3}).
We computed the P-values via 4,000,000 Monte-Carlo simulations
(i.e., 4,000,000 per empirical P-value being evaluated),
with each simulation taking $n = 20$ draws from the model.
The root-mean-square consistently and with extremely high confidence
rejects the hypothesis that the data arises from the model,
whereas the classical statistics find less and less evidence
for rejecting the hypothesis as the number $m$ of bins increases;
in fact, the P-values for the classical statistics get very close
to 1 as $m$ increases --- the discrepancy of~(\ref{synth3}) from 0
is usually less than the discrepancy of~(\ref{synth3})
from a typical realization drawn from the model,
since under the model the sum of the expected numbers
of draws in bins $3$,~$4$, \dots,~$m$\; is\; $n/2$.

Figure~\ref{plotsynth} demonstrates that the root-mean-square can be
much more powerful than the classical statistics, rejecting
with nearly 100\% confidence while the classical statistics
report nearly 0\% confidence for rejection.
Moreover, the classical statistics can report P-values
very close to 1 even when the data manifestly does not arise from the model.
(Incidentally, the model for smaller $m$ can be viewed as a rebinning
of the model for larger $m$. The classical statistics do reject the model
for smaller $m$, while asserting for larger $m$ that there is no evidence
for rejecting the model.)
The performance of the classical statistics displays a dramatic dependence
on the number ($m-2$) of unlikely bins in the model,
even though the data are the same for all $m$.
This suggests a sure-fire scheme for supporting any model
(no matter how invalid) with arbitrarily high P-values:
just append enough irrelevant, more or less uniformly improbable bins
to the model, and then report the P-values
for the classical goodness-of-fit statistics.
In contrast, the root-mean-square robustly and reliably rejects
the invalid model, independently of the size of the model.

We will see in the following section that
the classic Zipf power law behaves similarly.

For another example, we again consider the model specified
in~(\ref{synth1})--(\ref{synth3}).
For the empirical distribution, we now use $n = 96$ draws,
with 36 in the first bin, 12 in the second bin,
1~each for bins 3,~4, \dots,~50, and no draw in any other bin.
As before, this data clearly is unlikely to arise from the model
specified in~(\ref{synth1})--(\ref{synth3}), but we would like to see exactly
how well the various goodness-of-fit statistics detect the obvious discrepancy.

\begin{figure}
\begin{center}
\rotatebox{-90}{\scalebox{.47}{\includegraphics{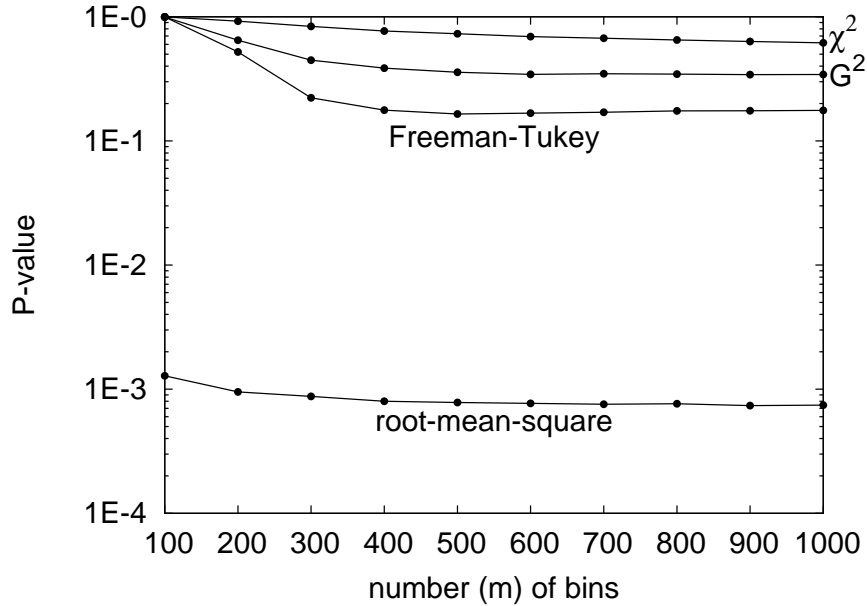}}}
\\\vspace{.1in}
\caption{P-values for the hypothesis
         that the model~(\ref{synth1})--(\ref{synth3}) agrees
         with the data of 36 draws in the first bin,
         12 draws in the second bin, 1 draw each in bins 3,~4, \dots,~50,
         and no draw in any other bin}
\label{plotstwo}
\end{center}
\end{figure}

Figure~\ref{plotstwo} plots the P-values for testing
whether the empirical data arises from the model
specified in~(\ref{synth1})--(\ref{synth3}).
We computed the P-values via 160,000 Monte-Carlo simulations
(that is, 160,000 per empirical P-value being evaluated),
with each simulation taking $n = 96$ draws from the model.
Yet again, the root-mean-square consistently and confidently
rejects the hypothesis that the data arises from the model,
whereas the classical statistics find little evidence for rejecting
the manifestly invalid model.

\subsection{Zipf's power law of word frequencies}
\label{zipfsec}

Zipf popularized his eponymous law by analyzing four
``chief sources of statistical data referred to in the main text''
(this is a quotation from the ``Notes and References'' section --- page~311 ---
of~\cite{zipf});
the chief source for the English language is~\cite{eldridge}.
We revisit the data of~\cite{eldridge} in the present subsection
to assess the performance of the goodness-of-fit statistics.

\begin{figure}
\begin{center}
\rotatebox{-90}{\scalebox{.47}{\includegraphics{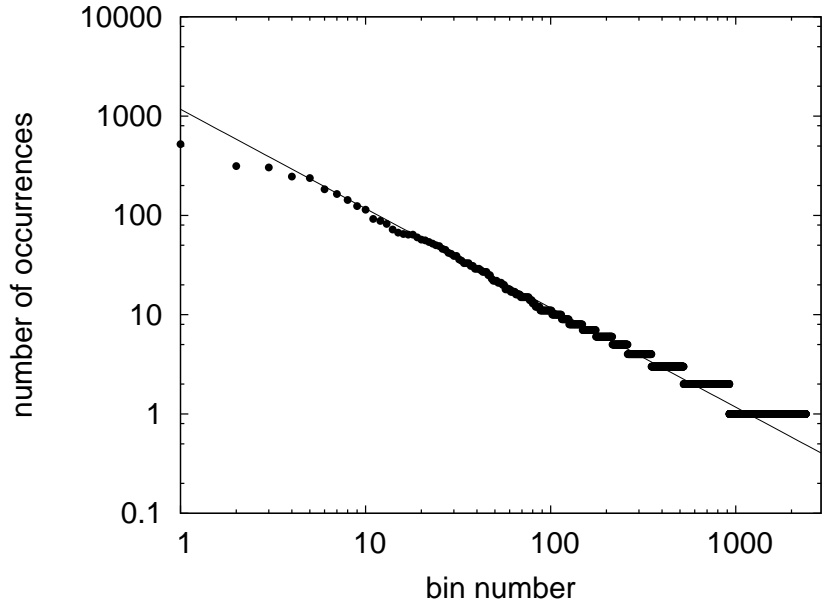}}}
\\\vspace{.1in}
\caption{Numbers of occurrences of the various words
         (one bin for each distinct word)
         in a corpus of 10,000 random draws from List~1 of~\cite{eldridge}}
\label{partdist}
\end{center}
\end{figure}

We first analyze List~1 of~\cite{eldridge},
which consists of 2,890 different English words,
such that there are 13,825 words in total counting repetitions;
the words come from the Buffalo Sunday News of August 8, 1909.
We randomly choose $n =$ 10,000 of the 13,825 words
to obtain a corpus of $n =$ 10,000 draws over 2,890 bins.
Figure~\ref{partdist} plots the frequencies
of the different words when sorted in rank order (so that the frequencies
are nonincreasing).
Using goodness-of-fit statistics we test the significance
of the (null) hypothesis that the empirical draws actually arise
from the Zipf distribution
\begin{equation}
\label{hypothetical}
p_0^{(j)}(\theta) = \frac{C_1}{\theta(j)}
\end{equation}
for $j = 1$,~$2$, \dots,~$m$, where
$\theta$ is a permutation of the integers $1$,~$2$, \dots,~$m$, and
\begin{equation}
\label{hconst}
C_1 = \frac{1}{\sum_{j=1}^m 1/j};
\end{equation}
we estimate the permutation $\theta$ via maximum-likelihood methods, that is,
by sorting the frequencies:
first we choose $j_1$ to be the number of a bin containing
the greatest number of draws among all $m$ bins,
then we choose $j_2$ to be the number of a bin containing
the greatest number of draws among the remaining $m-1$ bins,
then we choose $j_3$ to be the number of a bin containing
the greatest among the remaining $m-2$ bins, and so on,
and finally we find $\theta$ such that $\theta(j_1) = 1$,~$\theta(j_2) = 2$,
\dots,~$\theta(j_m) = m$.
We have to obtain the ordering $\theta$ from the data via such sorting
since we do not know the proper ordering a priori.

\begin{figure}
\begin{center}
\rotatebox{-90}{\scalebox{.47}{\includegraphics{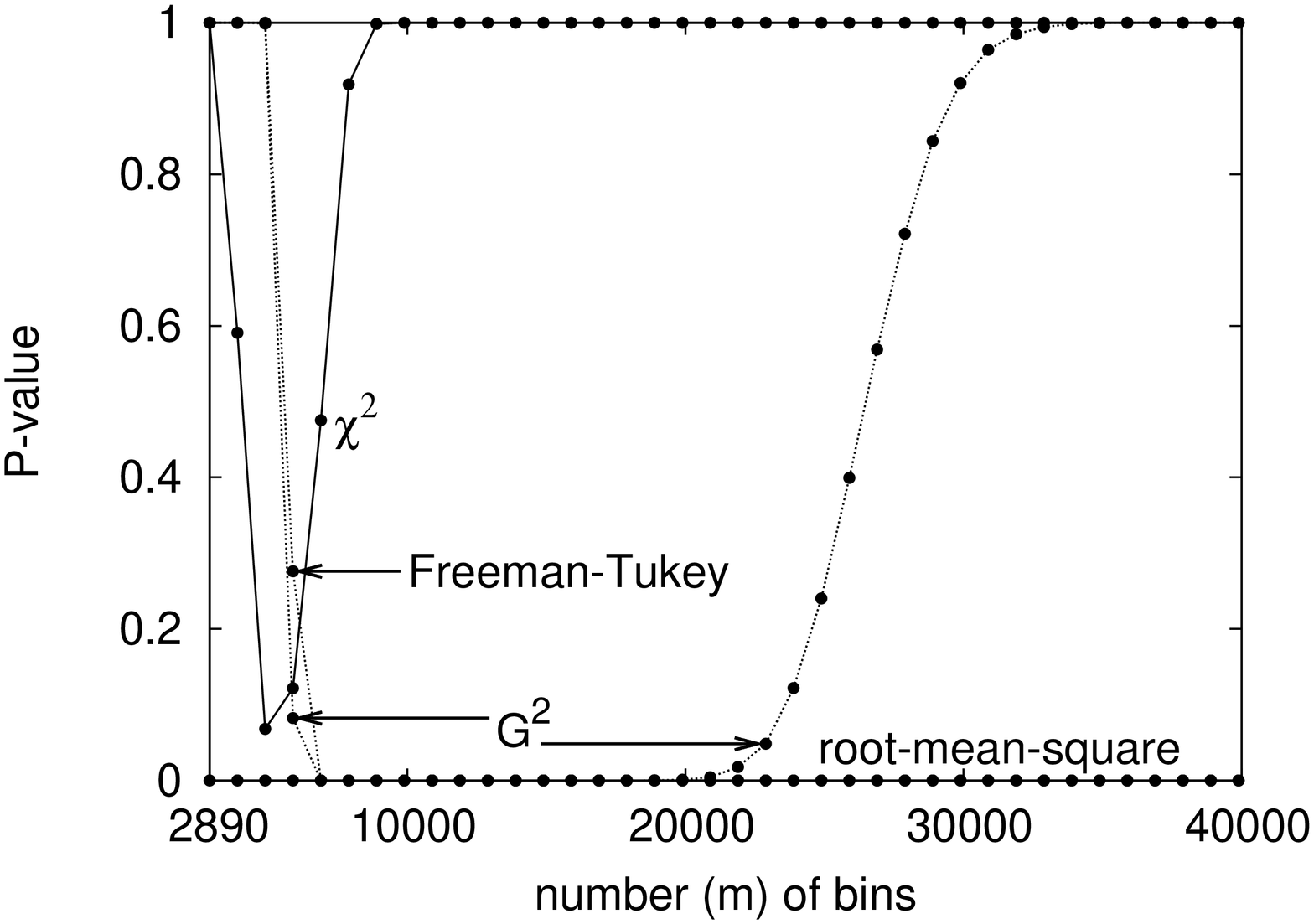}}}
\\\vspace{.1in}
\caption{P-values for the data plotted in Figure~\ref{partdist}
         to follow the Zipf distribution}
\label{partnopar}
\end{center}
\end{figure}

Similarly, we do not know the proper value of the number $m$ of bins,
so in Figure~\ref{partnopar} we plot P-values
(each computed via 40,000 Monte-Carlo simulations) for varying values of $m$;
although List~1 of~\cite{eldridge} involves only 2,890 distinct words,
we must also include bins for words that did not appear in the original list,
words whose frequencies are zeros for List~1 of~\cite{eldridge}.
Note that Figure~\ref{partnopar} displays the P-values
with $m =$ 2,890 for reference,
even though $m$ must be independent of the data,
and so $m$ must be substantially larger than 2,890
in order for the assumptions of goodness-of-fit testing to hold.

With respect to testing goodness-of-fit, the number $m$ of bins
is the number of words in the dictionary
from which List~1 of~\cite{eldridge} was drawn.
It is not clear a priori which dictionary is appropriate.
Fortunately, the P-values for the root-mean-square
are always 0 to several digits of accuracy,
independent of the value of $m$ --- the root-mean-square determines
that List~1 does not follow the classic Zipf distribution
(defined in~(\ref{hypothetical}) and~(\ref{hconst})) for any $m$.
In contrast, the P-values for the classical statistics
vary wildly depending on the value of $m$.
In fact, for any of the classical statistics,
and for any prescribed number $P$ between 0.05 and 0.95,
there is at least one value of $m$ between 4,000 and 40,000
such that the P-value is $P$.
Thus, without knowing the proper size of the dictionary a priori,
the classical statistics are meaningless.

\begin{figure}
\begin{center}
\rotatebox{-90}{\scalebox{.47}{\includegraphics{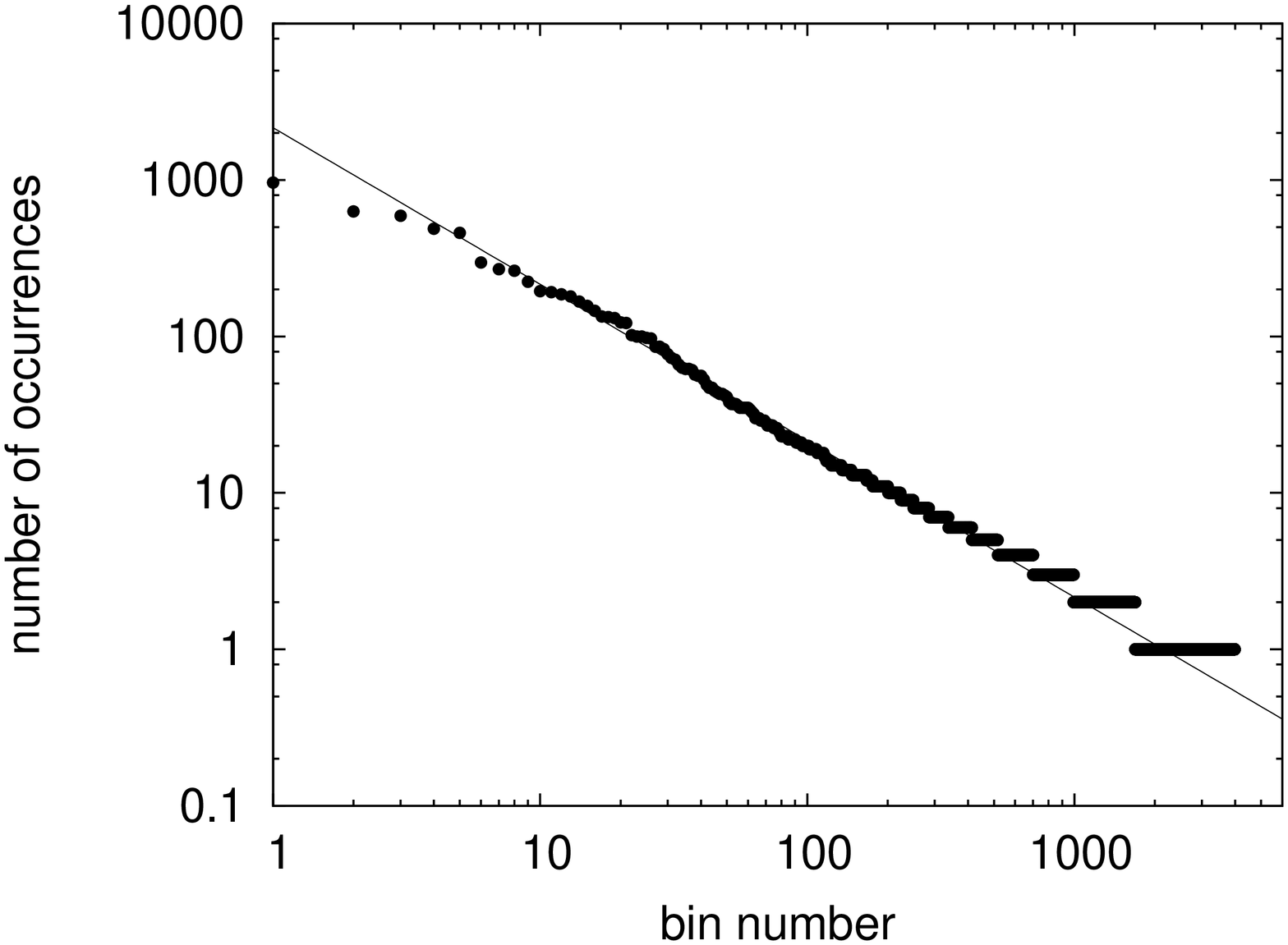}}}
\\\vspace{.1in}
\caption{Numbers of occurrences of the various words
         (one bin for each distinct word)
         in a corpus of 20,000 random draws from List~5 of~\cite{eldridge}}
\label{fulldist}
\end{center}
\end{figure}

Unsurprisingly, analyzing List~5 of~\cite{eldridge}
produces results analogous to those reported above for List~1.
List~5 consists of 6,002 different English words,
such that there are 43,989 words in total counting repetitions;
the words come from amalgamating Lists~1--4 of~\cite{eldridge}.
We randomly choose $n =$ 20,000 of the 43,989 words
to obtain a corpus of $n =$ 20,000 draws over 6,002 bins.
Figure~\ref{fulldist} plots the frequencies
of the different words when sorted in rank order (so that the frequencies
are nonincreasing).

\begin{figure}
\begin{center}
\rotatebox{-90}{\scalebox{.47}{\includegraphics{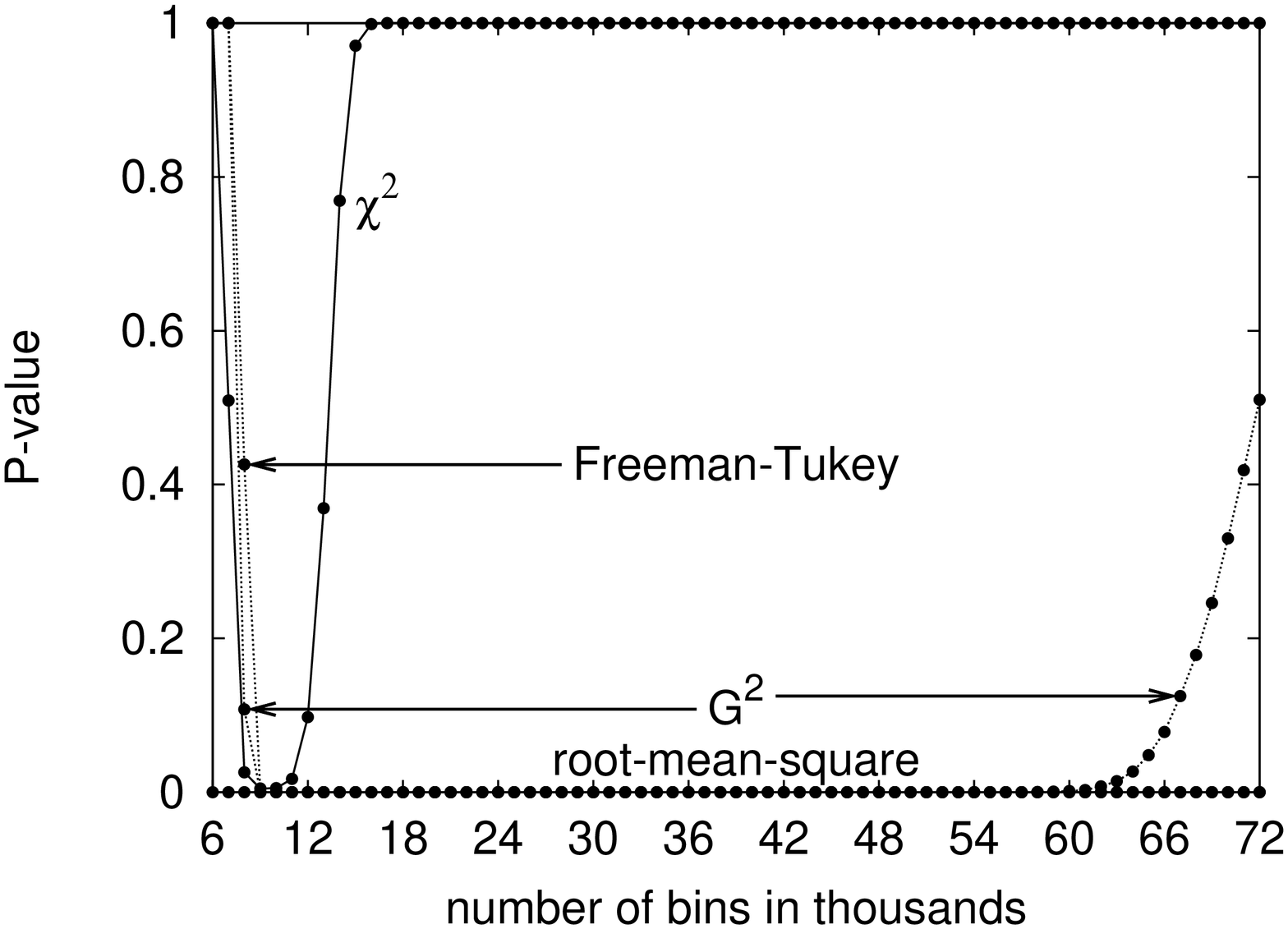}}}
\\\vspace{.1in}
\caption{P-values for the data plotted in Figure~\ref{fulldist}
         to follow the Zipf distribution}
\label{fullnopar}
\end{center}
\end{figure}

Again we do not know the proper value of the number $m$ of bins,
so in Figure~\ref{fullnopar} we plot P-values
(each computed via 40,000 Monte-Carlo simulations) for varying values of $m$;
although List~5 of~\cite{eldridge} involves only 6,002 distinct words,
we must also include bins for words that did not appear in the original list,
words whose frequencies are zeros for List~5 of~\cite{eldridge}.
Please note that Figure~\ref{fullnopar} displays the P-values
with $m =$ 6,002 for reference,
even though $m$ must be independent of the data,
and so $m$ must be substantially larger than 6,002
in order for the assumptions of goodness-of-fit testing to hold.
Comparing Figures~\ref{partnopar} and~\ref{fullnopar} shows that
the above remarks about List~1 pertain to the analysis of the larger List~5,
too. Once again, without knowing the proper size of the dictionary a priori,
the classical statistics are meaningless, whereas the root-mean-square is
very powerful.

Interestingly,
by introducing parameters $\theta_1$, $\theta_2$, and $\theta_3$
to fit perfectly the bins containing the three greatest numbers of draws,
a truncated power-law becomes a good fit
for the corpus of 20,000 words drawn randomly from List~5 of~\cite{eldridge},
with the number $m$ of bins set to 7,500.
Indeed, let us consider the model
\begin{equation}
\label{complicated1}
p_0^{(j)}(\theta_0,\theta_1,\theta_2,\theta_3,\theta_4)
= \left\{ \begin{array}{ll}
          \theta_1, & \theta_0(j) = 1 \\
          \theta_2, & \theta_0(j) = 2 \\
          \theta_3, & \theta_0(j) = 3 \\
          C/(\theta_0(j))^{\theta_4},
          & \theta_0(j) = 4, 5, \dots, 7500
          \end{array} \right.,
\end{equation}
where
\begin{equation}
\label{complicated2}
C = C_{\theta_1,\theta_2,\theta_3,\theta_4}
  = \frac{1-\theta_1-\theta_2-\theta_3}{\sum_{j=4}^{7500} 1/j^{\theta_4}},
\end{equation}
with $\theta_0$ being a permutation
of the integers $1$,~$2$, \dots,~$7500$,
and $\theta_1$, $\theta_2$, $\theta_3$, $\theta_4$
being nonnegative real numbers;
we estimate $\theta_0$, $\theta_1$, $\theta_2$, $\theta_3$, $\theta_4$
via maximum-likelihood methods,
determining $\theta_0$ by sorting as discussed above, and
setting $\theta_1$, $\theta_2$, and~$\theta_3$
to be the three greatest relative frequencies.
This model fits the empirical data exactly
in the bins whose probabilities under the model
are $\theta_1$, $\theta_2$, and $\theta_3$ ---
there will be no discrepancy between the data and the model in those bins ---
so that these bins do not contribute to any goodness-of-fit statistic,
aside from altering the number of draws in the remaining bins.
Of the 20,000 total draws in the given experimental data,
16,486 do not fall in the bins
associated with the three most frequently occurring words.
The maximum-likelihood estimate of the power-law exponent $\theta_4$
for the experimental data turns out to be about 1.0484.

For the model defined in~(\ref{complicated1}) and~(\ref{complicated2}),
the P-values calculated via 4,000,000 Monte-Carlo simulations are
\begin{itemize}
\item $\chi^2$: .510
\item $G^2$: .998
\item Freeman-Tukey: 1.000
\item root-mean-square: .587
\end{itemize}
Thus, all four statistics indicate that the truncated power-law model
defined in~(\ref{complicated1}) and~(\ref{complicated2}) is a good fit.
This is in accord with Figure~\ref{fulldist},
in which all but the three greatest frequencies appear
to follow a truncated power-law.

\subsection{A Poisson law for radioactive decays}

\begin{figure}
\begin{center}
\rotatebox{-90}{\scalebox{.47}{\includegraphics{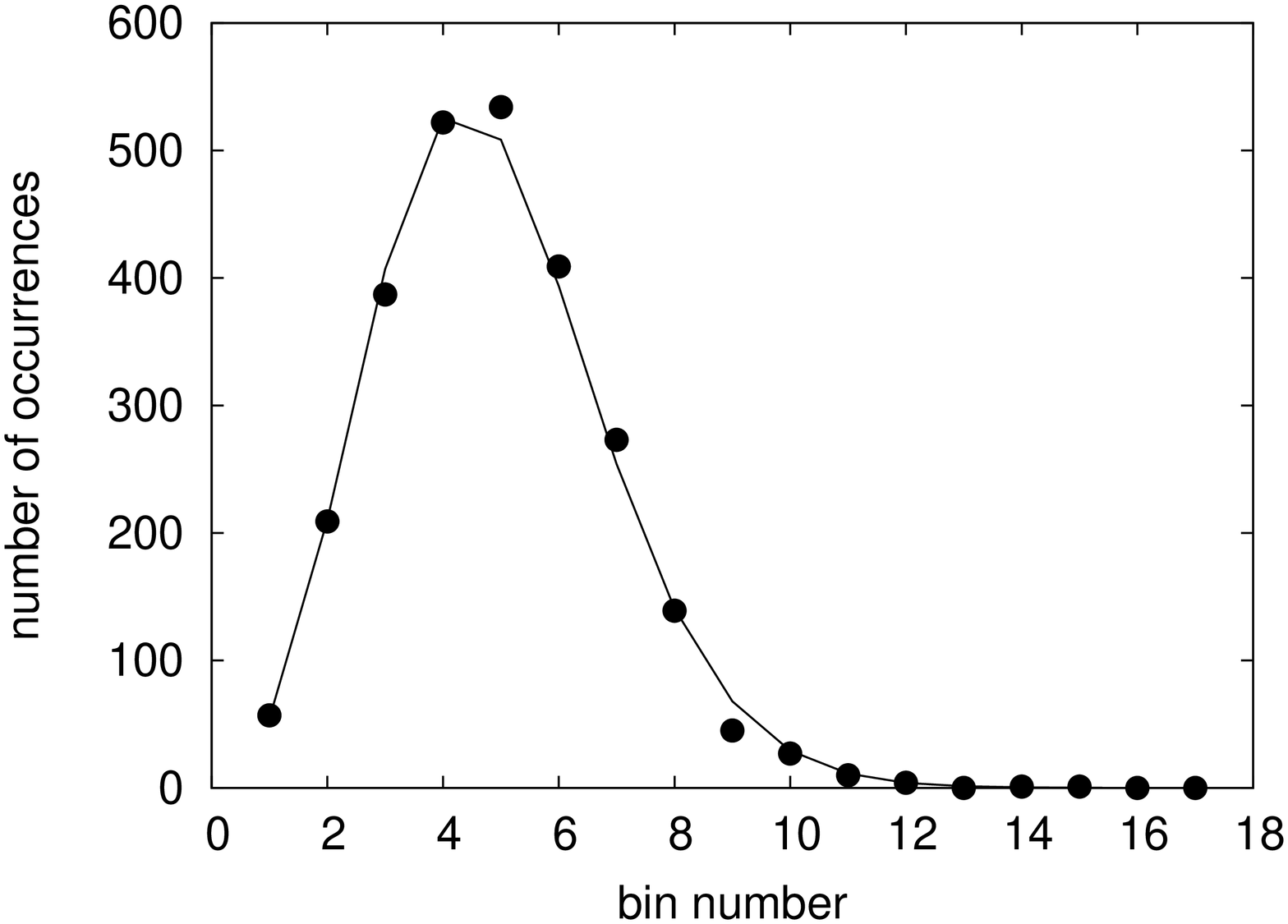}}}
\\\vspace{.1in}
\caption{The data in Table~\ref{poissont} (the dots) and
         the best-fit Poisson distribution (the lines)}
\label{alphadist}
\end{center}
\end{figure}

\begin{figure}
\begin{center}
\rotatebox{-90}{\scalebox{.47}{\includegraphics{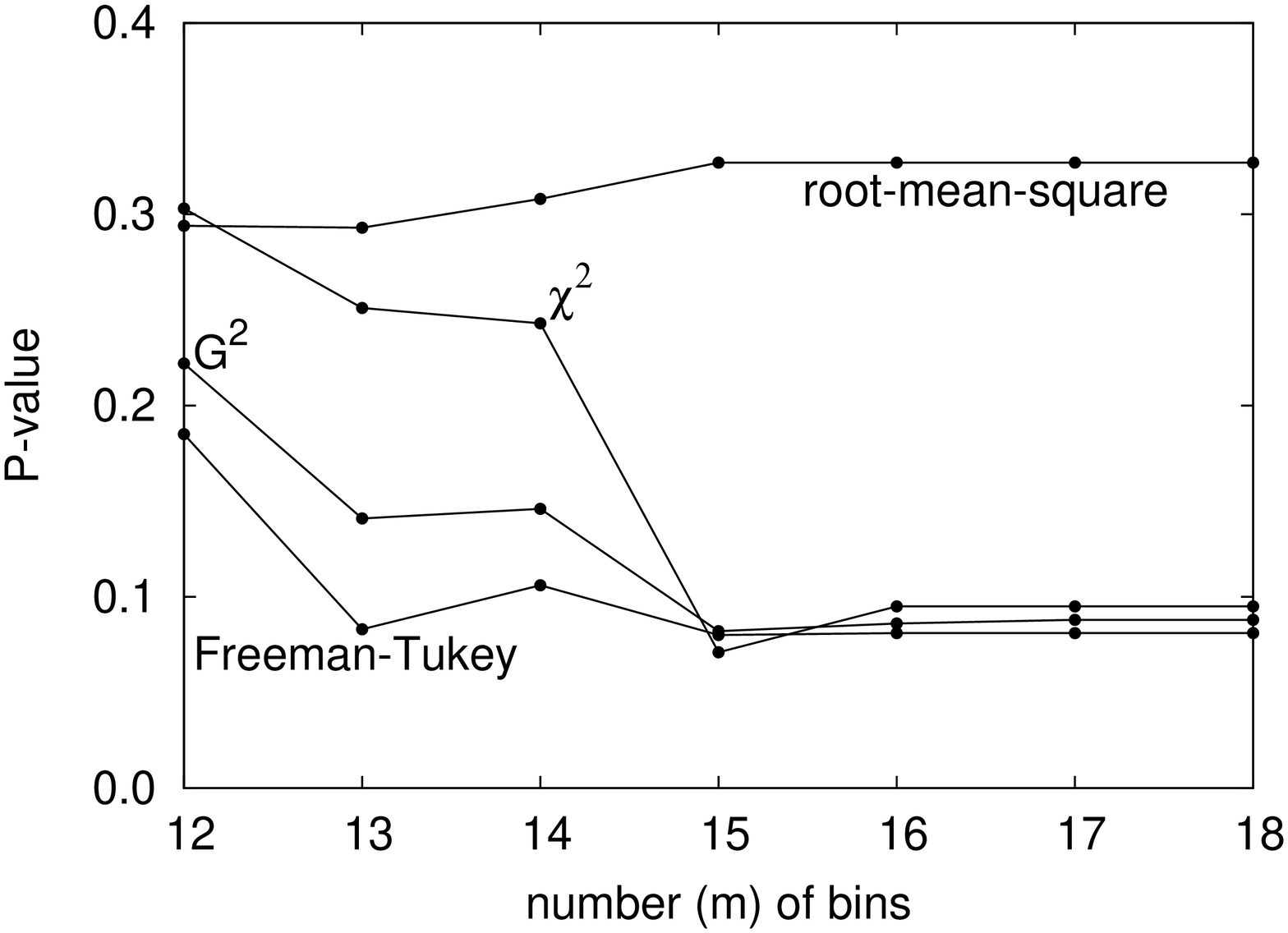}}}
\\\vspace{.1in}
\caption{P-values for the distribution of Table~\ref{poissont}
         to be Poisson}
\label{poissonpar}
\end{center}
\end{figure}

Table~\ref{poissont} summarizes the classic example
of a Poisson-distributed experiment in radioactive decay
of~\cite{rutherford-geiger-bateman};
Figure~\ref{alphadist} plots the data, along with the Poisson distribution
whose mean is the same as the data's.
Figure~\ref{poissonpar} reports the P-values
for testing whether the data,
while retaining only bins $1$,~$2$, \dots,~$m$,
are distributed according to a Poisson distribution
(the model Poisson distribution is also truncated to the first $m$ bins,
with the mean estimated from the data).
Since the total number $n$ of draws depends little on the numbers
in bins 13,~14, 15,~\dots, the truncation amounts to ignoring draws
in bins $m+1$,~$m+2$, $m+3$,~\dots\ when $m \ge 12$,
and demonstrates that the scant experimental draws in bins 13--15
strongly influence the P-values of the classical statistics.
We computed the P-values via 40,000 Monte-Carlo simulations
(for each number $m$ of bins and each of the four statistics),
estimating the mean of the model Poisson distribution
for each simulated data set.
All four goodness-of-fit statistics indicate reasonably good agreement
between the data and a Poisson distribution;
the classical statistics are very sensitive in the tail to discrepancies
between the data and the model distribution,
whereas the root-mean-square is relatively insensitive to the truncation
after 12 or more bins.

\begin{table}
\caption{Numbers of $\alpha$-particles emitted by a film of polonium
         in 2608 intervals of 7.5 seconds}
\label{poissont}
\begin{center}
\begin{tabular}{ccc}
& number of particles observed & \\
bin number & in an interval of 7.5 seconds & number of such intervals \\\hline
 1 &  0 &  57 \\
 2 &  1 & 203 \\
 3 &  2 & 383 \\
 4 &  3 & 525 \\
 5 &  4 & 532 \\
 6 &  5 & 408 \\
 7 &  6 & 273 \\
 8 &  7 & 139 \\
 9 &  8 &  45 \\
10 &  9 &  27 \\
11 & 10 &  10 \\
12 & 11 &   4 \\
13 & 12 &   0 \\
14 & 13 &   1 \\
15 & 14 &   1 \\
16, 17, 18, \dots & 15, 16, 17, \dots & 0 \\\hline
1, 2, 3, 4, 5, \dots & 0, 1, 2, 3, 4, \dots & 2608
\end{tabular}
\end{center}
\end{table}

\subsection{A Poisson law for counting with a h\ae{}macytometer}

\begin{figure}
\begin{center}
\rotatebox{-90}{\scalebox{.47}{\includegraphics{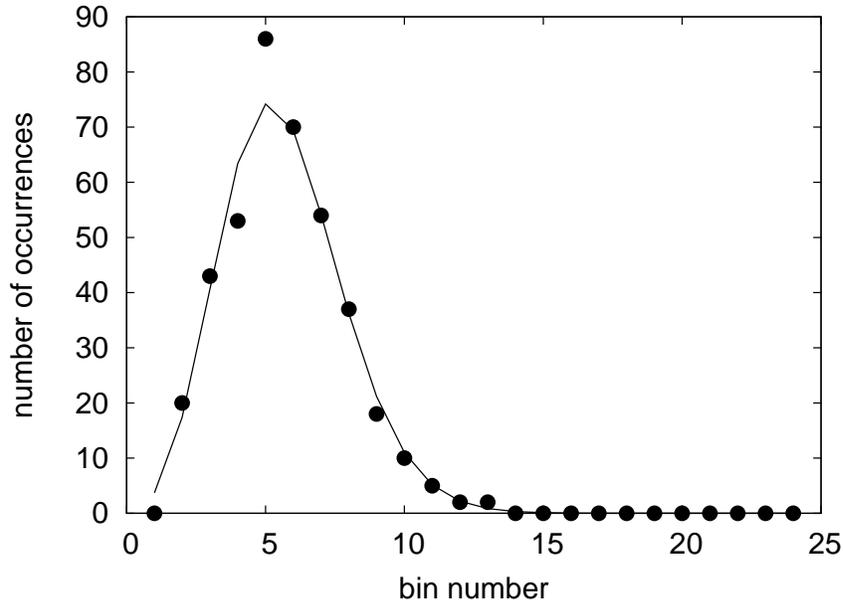}}}
\\\vspace{.1in}
\caption{The data in Table~\ref{yeast} (the dots) and
         the best-fit Poisson distribution (the lines)}
\label{yeastdist}
\end{center}
\end{figure}

Page~357 of~\cite{student} reports on the number of yeast cells
observed in each of 400 squares in a h\ae{}macytometer microscope slide.
Table~\ref{yeast} displays the counts;
Figure~\ref{yeastdist} plots them, along with the Poisson distribution
whose mean matches the data's.
The P-values for the data to arise from a Poisson distribution
(with the mean estimated from the data) are

\begin{itemize}
\item $\chi^2$: .627
\item $G^2$: .365
\item Freeman-Tukey: .111
\item root-mean-square: .490
\end{itemize}
We calculated the P-values via 4,000,000 Monte-Carlo simulations,
estimating the mean of the model Poisson distribution
for each simulated data set.
Evidently, all four statistics report that a Poisson distribution
is a reasonably good model for the experimental data.

\begin{table}
\caption{Numbers of yeast cells in 400 squares of a h\ae{}macytometer}
\label{yeast}
\begin{center}
\begin{tabular}{ccc}
bin number & number of yeast in a square & number of such squares \\\hline
 1 &  0 &  0 \\
 2 &  1 & 20 \\
 3 &  2 & 43 \\
 4 &  3 & 53 \\
 5 &  4 & 86 \\
 6 &  5 & 70 \\
 7 &  6 & 54 \\
 8 &  7 & 37 \\
 9 &  8 & 18 \\
10 &  9 & 10 \\
11 & 10 &  5 \\
12 & 11 &  2 \\
13 & 12 &  2 \\
14, 15, 16, \dots & 13, 14, 15, \dots & 0 \\\hline
1, 2, 3, 4, 5, \dots & 0, 1, 2, 3, 4, \dots & 400
\end{tabular}
\end{center}
\end{table}

\subsection{A Hardy-Weinberg law for Rhesus blood groups}
\label{hardyweinberg}

In a population with suitably random mating,
the proportions of pairs of Rhesus haplotypes in members of the population
(each member has one pair) can be expected to follow the Hardy-Weinberg law
discussed by~\cite{guo-thompson},
namely to arise via random sampling from the model
\begin{equation}
\label{hw}
p_0^{(j,k)}(\theta_1, \theta_2, \dots, \theta_9)
= \left\{ \begin{array}{ll}
          2 \cdot \theta_j \cdot \theta_k, & j > k \\
          (\theta_k)^2, & j = k
  \end{array} \right.
\end{equation}
for $j,k = 1$,~$2$, \dots,~$9$ with $j \ge k$, under the constraint that
\begin{equation}
\sum_{j=1}^9 \theta_j = 1,
\end{equation}
where the parameters $\theta_1$,~$\theta_2$, \dots,~$\theta_9$
are the proportions of the nine Rhesus haplotypes in the population
(naturally, their maximum-likelihood estimates are the proportions
of the haplotypes in the given data).
For $j,k = 1$,~$2$, \dots,~$9$ with $j \ge k$, therefore,
$p_0^{(j,k)}$ is the expected probability that the pair of haplotypes
in the genome of an individual is the pair $j$ and $k$,
given the parameters $\theta_1$,~$\theta_2$, \dots,~$\theta_9$.

In this formulation, the hypothesis of suitably random mating entails that
the members of the sample population are i.i.d.\ draws from the model specified
in~(\ref{hw}); if a goodness-of-fit statistic rejects the model
with high confidence, then we can be confident that mating
has not been suitably random.
Table~\ref{hwt} provides data on $n = 8297$ individuals;
we duplicated Figure~3 of~\cite{guo-thompson} to obtain Table~\ref{hwt}.

\begin{table}
\caption{Frequencies of pairs of Rhesus haplotypes}
\label{hwt}
\begin{center}
\hspace{3.5pc}$k$\\\vspace{4pt}
$j$
\begin{tabular}{c||c|c|c|c|c|c|c|c|c}
\hspace{-1pc}$_{j\hspace{-.3pc}}\diagdown{}^{\hspace{-.35pc}k}$\hspace{-1.3pc}
& 1 & 2 & 3 & 4 & 5 & 6 & 7 & 8 & 9 \\
\hline\hline
1 & 1236 &&&&&&& \\\hline
2 & 120 & 3 &&&&&&& \\\hline
3 & 18 & 0 & 0 &&&&& \\\hline
4 & 982 & 55 & 7 & 249 &&&& \\\hline
5 & 32 & 1 & 0 & 12 & 0 &&& \\\hline
6 & 2582 & 132 & 20 & 1162 & 29 & 1312 && \\\hline
7 & 6 & 0 & 0 & 4 & 0 & 4 & 0 & \\\hline
8 & 2 & 0 & 0 & 0 & 0 & 0 & 0 & 0 \\\hline
9 & 115 & 5 & 2 & 53 & 1 & 149 & 0 & 0 & 4
\end{tabular}
\end{center}
\end{table}

The P-values calculated via 4,000,000 Monte-Carlo simulations are

\begin{itemize}
\item $\chi^2$: .693
\item $G^2$: .600
\item Freeman-Tukey: .562
\item negative log-likelihood (see Remark~\ref{nll} below): .649
\item root-mean-square: .039
\end{itemize}
Unlike the root-mean-square, the classical statistics are blind
to the significant discrepancy between the data and the Hardy-Weinberg model.
Please note that the P-values associated with the classical statistics
are over an order of magnitude larger than the P-value associated
with the root-mean-square.

\begin{remark}
For the example of the present subsection,
rejecting the null hypothesis~(\ref{null'}) from Section~\ref{hypotheses}
might seem in principle to be more interesting
than rejecting the assumption~(\ref{null}).
Fortunately, the difference between~(\ref{null'}) and~(\ref{null})
is essentially irrelevant for the root-mean-square in this example.
Indeed, the root-mean-square is not very sensitive to bins associated
with the parameters whose estimated values are potentially inaccurate ---
the potentially inaccurate estimates are all small,
and the root-mean-square is not very sensitive
to bins whose probabilities under the model are small relative to others.
\end{remark}

\begin{remark}
\label{nll}
The term ``negative log-likelihood'' used in the present section refers
to the statistic that is simply the negative of the logarithm
of the likelihood.
The negative log-likelihood is the same statistic used
in the generalization of Fisher's exact test
discussed by~\cite{guo-thompson}; unlike $G^2$,
this statistic involves only one likelihood, not the ratio of two.
We mention the negative log-likelihood just to facilitate comparisons;
we are not asserting that the likelihood
on its own (rather than in a ratio) is a good gauge
of the relative sizes of deviations from a model.
\end{remark}

\begin{remark}
Table~\ref{hwt2} provides data on $n = 45$ individuals
from the other set of real-world measurements given by~\cite{guo-thompson};
we duplicated Figure~2 of~\cite{guo-thompson} to obtain Table~\ref{hwt2}.
The associated Hardy-Weinberg model is then the same as~(\ref{hw}),
but with only four parameters, $\theta_1$,~$\theta_2$,~$\theta_3$,~$\theta_4$,
such that $\sum_{j=1}^4 \theta_j = 1$.
The P-values calculated via 4,000,000 Monte-Carlo simulations are
\begin{itemize}
\item $\chi^2$: .021
\item $G^2$: .013
\item Freeman-Tukey: .027
\item negative log-likelihood (see Remark~\ref{nll} above): .016
\item root-mean-square: .0019
\end{itemize}
Again the root-mean-square is more powerful than the classical statistics.
\end{remark}

\begin{table}
\caption{Frequencies of genotypes}
\label{hwt2}
\begin{center}
\hspace{3.5pc}$k$\\\vspace{4pt}
\begin{tabular}{c}\\$j$\end{tabular}\hspace{-.5pc}
\begin{tabular}{c||c|c|c|c}
\hspace{-1pc}$_{j\hspace{-.3pc}}\diagdown{}^{\hspace{-.35pc}k}$\hspace{-1.3pc}
& 1 & 2 & 3 & 4 \\
\hline\hline
1 & 0 &&& \\\hline
2 & 3 & 1 && \\\hline
3 & 5 & 18 & 1 & \\\hline
4 & 3 & 7 & 5 & 2
\end{tabular}
\end{center}
\end{table}

\subsection{Symmetry between the self-reported health assessments
            of foreign- and US-born Asian Americans}

Using propensity scores, \cite{erosheva-walton-takeuchi} matched
each of 335 surveyed foreign-born Asian Americans to a similar
surveyed US-born Asian American.
Table~\ref{health} duplicates Table~4 of~\cite{erosheva-walton-takeuchi},
tabulating the numbers of matched pairs reporting various combinations
of physical health; the propensity scores were generated
without reference to the health ratings.
Table~\ref{health} does not reveal any significant difference
between foreign-born Asian Americans' ratings of their health
and US-born Asian Americans'.
Indeed, the P-values calculated via 4,000,000 Monte-Carlo simulations
for testing the symmetry of Table~\ref{health} are
\begin{itemize}
\item $\chi^2$: .784
\item $G^2$: .739
\item Freeman-Tukey: .642
\item root-mean-square: .973
\end{itemize}

After noting that $\chi^2$ does not reveal any statistically significant
asymmetry in Table~\ref{health},
\cite{erosheva-walton-takeuchi} reported that,
``to address the issue of power of this test,
we investigated what is the smallest departure from symmetry
that our test could detect\dots.''
Such an investigation requires considering modifications to Table~\ref{health}.
Table~\ref{healthmod} provides one possible modification.
The P-values calculated via 4,000,000 Monte-Carlo simulations
for testing the symmetry of Table~\ref{healthmod} are
\begin{itemize}
\item $\chi^2$: .109
\item $G^2$: .123
\item Freeman-Tukey: .155
\item root-mean-square: .014
\end{itemize}
Evidently, the root-mean-square is more powerful for detecting the asymmetry
of Table~\ref{healthmod}.

Table~\ref{healthmods} provides another hypothetical cross-tabulation.
The P-values calculated via 64,000,000 Monte-Carlo simulations
for testing the symmetry of Table~\ref{healthmods} are
\begin{itemize}
\item $\chi^2$: .0015
\item $G^2$: .00016
\item Freeman-Tukey: .000006, i.e., 6E--6
\item root-mean-square: .131
\end{itemize}
The classical statistics are much more powerful for detecting the asymmetry
of Table~\ref{healthmods}, contrasting how the root-mean-square
is more powerful for detecting the asymmetry of Table~\ref{healthmod}.
Indeed, the root-mean-square statistic is not very sensitive
to relative discrepancies between the model and actual distributions
in bins whose associated model probabilities are small.
When sensitivity in these bins is desirable,
we recommend using both the root-mean-square statistic
and an asymptotically equivalent variation of $\chi^2$
such as the log--likelihood-ratio $G^2$.

\begin{table}[p]
\caption{Self-reported physical health for matched pairs of Asian Americans}
\label{health}
\begin{center}
\begin{tabular}{cc|ccccc}
&&&&foreign-born\\&&&\\
&& excellent & very good & good & fair & poor \\\hline
& excellent
& \ \ \ \quad 10 \quad\ \ \ & \ \ \ \quad 21 \quad\ \ \ 
& \ \ \ \quad 22 \quad\ \ \ & \ \ \ \quad  5 \quad\ \ \ 
& \ \ \ \quad 0 \quad\ \ \ \\
& very good
& \ \ \ \quad 24 \quad\ \ \ & \ \ \ \quad 53 \quad\ \ \ 
& \ \ \ \quad 43 \quad\ \ \ & \ \ \ \quad 15 \quad\ \ \ 
& \ \ \ \quad  3 \quad\ \ \ \\
US-born & good
& \ \ \ \quad 21 \quad\ \ \ & \ \ \ \quad 43 \quad\ \ \ 
& \ \ \ \quad 34 \quad\ \ \ & \ \ \ \quad 11 \quad\ \ \ 
& \ \ \ \quad  0 \quad\ \ \ \\
& fair
& \ \ \ \quad  3 \quad\ \ \ & \ \ \ \quad 11 \quad\ \ \ 
& \ \ \ \quad  8 \quad\ \ \ & \ \ \ \quad  4 \quad\ \ \ 
& \ \ \ \quad  1 \quad\ \ \ \\
& poor
& \ \ \ \quad  1 \quad\ \ \ & \ \ \ \quad  1 \quad\ \ \ 
& \ \ \ \quad  1 \quad\ \ \ & \ \ \ \quad  0 \quad\ \ \ 
& \ \ \ \quad  0 \quad\ \ \ 
\end{tabular}
\end{center}
\end{table}

\begin{table}
\caption{A variation on Table~\ref{health}}
\label{healthmod}
\begin{center}
\begin{tabular}{cc|ccccc}
&&&&foreign-born\\&&&\\
&& excellent & very good & good & fair & poor \\\hline
& excellent
& \ \ \ \quad 10 \quad\ \ \ & \ \ \ \quad 21 \quad\ \ \ 
& \ \ \ \quad 22 \quad\ \ \ & \ \ \ \quad  5 \quad\ \ \ 
& \ \ \ \quad 0 \quad\ \ \ \\
& very good
& \ \ \ \quad 24 \quad\ \ \ & \ \ \ \quad 53 \quad\ \ \ 
& \ \ \ \quad {\it 56} \quad\ \ \ & \ \ \ \quad 15 \quad\ \ \ 
& \ \ \ \quad  3 \quad\ \ \ \\
US-born & good
& \ \ \ \quad 21 \quad\ \ \ & \ \ \ \quad {\it 30} \quad\ \ \ 
& \ \ \ \quad 34 \quad\ \ \ & \ \ \ \quad 11 \quad\ \ \ 
& \ \ \ \quad  0 \quad\ \ \ \\
& fair
& \ \ \ \quad  3 \quad\ \ \ & \ \ \ \quad 11 \quad\ \ \ 
& \ \ \ \quad  8 \quad\ \ \ & \ \ \ \quad  4 \quad\ \ \ 
& \ \ \ \quad  1 \quad\ \ \ \\
& poor
& \ \ \ \quad  1 \quad\ \ \ & \ \ \ \quad  1 \quad\ \ \ 
& \ \ \ \quad  1 \quad\ \ \ & \ \ \ \quad  0 \quad\ \ \ 
& \ \ \ \quad  0 \quad\ \ \ 
\end{tabular}
\end{center}
\end{table}

\begin{table}
\caption{Another variation on Table~\ref{health}}
\label{healthmods}
\begin{center}
\begin{tabular}{cc|ccccc}
&&&&foreign-born\\&&&\\
&& excellent & very good & good & fair & poor \\\hline
& excellent
& \ \ \ \quad 10 \quad\ \ \ & \ \ \ \quad 21 \quad\ \ \ 
& \ \ \ \quad 22 \quad\ \ \ & \ \ \ \quad  5 \quad\ \ \ 
& \ \ \ \quad 0 \quad\ \ \ \\
& very good
& \ \ \ \quad 24 \quad\ \ \ & \ \ \ \quad 53 \quad\ \ \ 
& \ \ \ \quad 43 \quad\ \ \ & \ \ \ \quad 15 \quad\ \ \ 
& \ \ \ \quad  3 \quad\ \ \ \\
US-born & good
& \ \ \ \quad 21 \quad\ \ \ & \ \ \ \quad 43 \quad\ \ \ 
& \ \ \ \quad 34 \quad\ \ \ & \ \ \ \quad {\it 19} \quad\ \ \ 
& \ \ \ \quad  0 \quad\ \ \ \\
& fair
& \ \ \ \quad  3 \quad\ \ \ & \ \ \ \quad 11 \quad\ \ \ 
& \ \ \ \quad  {\it 0} \quad\ \ \ & \ \ \ \quad  4 \quad\ \ \ 
& \ \ \ \quad  1 \quad\ \ \ \\
& poor
& \ \ \ \quad  1 \quad\ \ \ & \ \ \ \quad  1 \quad\ \ \ 
& \ \ \ \quad  1 \quad\ \ \ & \ \ \ \quad  0 \quad\ \ \ 
& \ \ \ \quad  0 \quad\ \ \ 
\end{tabular}
\end{center}
\end{table}

\subsection{A modified geometric law for the species of butterflies}

\begin{figure}
\begin{center}
\rotatebox{-90}{\scalebox{.47}{\includegraphics{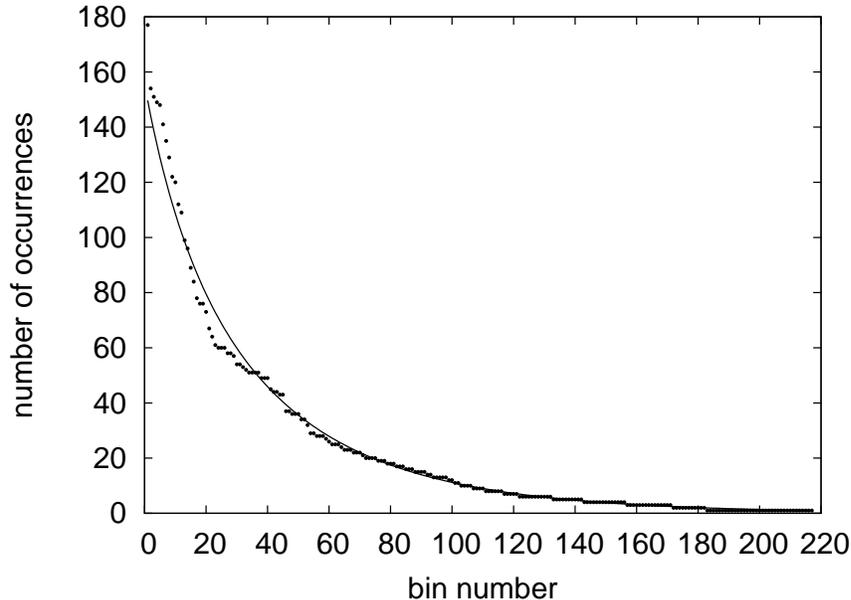}}}
\\\vspace{.1in}
\caption{Numbers of specimens (the dots) from
         217 species of butterflies (one bin per species), and
         the best-fit distribution (the lines)}
\label{butdist}
\end{center}
\end{figure}

\cite{fisher-corbet-williams} reported on 5300 butterflies
from 217 readily identified species
(these exclude the 23 most common readily identified species)
that they collected via random sampling
at the Rothamsted Experimental Station in England.
Figure~\ref{butdist} plots the numbers of individual butterflies collected
from the 217 species when sorted in rank order
(so that the numbers are nonincreasing).

To build a model appropriate for Figure~\ref{butdist},
we must include a permutation of the bins as a parameter,
since we have sorted the data (see Subsection~\ref{zipfsec}
for further discussion of sorting and permutations).
We take the model to be
\begin{equation}
\label{negbinom}
p_0^{(j)}(\theta_0, \theta_1)
= A_{\theta_1} \frac{(\theta_1)^{\theta_0(j)}}{\sqrt{\theta_0(j)+23}}
\end{equation}
for $j = 1$,~$2$, \dots,~$217$,
where $\theta_0$ is a permutation of the integers
$1$,~$2$, \dots,~$217$,
the parameter $\theta_1$ is a positive real number less than 1, and
\begin{equation}
A_{\theta_1}
= \frac{1}{\sum_{j=1}^{217} (\theta_1)^j/\sqrt{j+23}};
\end{equation}
we estimate $\theta_0$ and $\theta_1$ via maximum-likelihood methods
(thus obtaining $\theta_0$ by sorting the frequencies
into nonincreasing order).
Please note that this model is not very carefully chosen ---
the model is just a truncated geometric distribution
weighted by the nonsingular function $1/\sqrt{\theta_0(j)+23}$,
with 23 being the number of common species omitted from the collection.
More complicated models may fit better.

The P-values calculated via 4,000,000 Monte-Carlo simulations are
\begin{itemize}
\item $\chi^2$: .0050
\item $G^2$: .349
\item Freeman-Tukey: .951
\item root-mean-square: .00002, i.e., 2E--5
\end{itemize}
As Figure~\ref{butdist} indicates, the discrepancy between the empirical data
and the model is substantial, and, given the large number of draws (5300),
cannot be due solely to random fluctuations.
The log--likelihood-ratio ($G^2$) and Freeman-Tukey statistics
are unable to detect this discrepancy,
while the root-mean-square easily determines that the discrepancy
is very highly significant.

\section{The power and efficiency of the root-mean-square}
\label{power}

In this section, we consider many numerical experiments and models,
plotting the numbers of draws required for goodness-of-fit statistics
to detect divergence from the models.
We consider both fully specified models and parameterized models.
To quantify a statistic's success at detecting discrepancies from the models,
we use the formulation of the following remark.

\begin{remark}
\label{distinguish}
We say that a statistic based on given i.i.d.\ draws
``distinguishes'' the actual underlying distribution of the draws
from the model distribution to mean that the computed P-value is
at most {\it 1\%} for {\it 99\%} of 40,000 simulations,
with each simulation generating $n$ i.i.d.\ draws according
to the actual distribution.
We computed the P-values by conducting another 40,000 simulations,
with each simulation generating $n$ i.i.d.\ draws according
to the model distribution.
Appendix~A of~\cite{perkins-tygert-ward3}
uses a weaker notion of ``distinguish'' ---
in Appendix~A we say that a statistic based on given i.i.d.\ draws
``distinguishes'' the actual underlying distribution of the draws
from the model distribution to mean that the computed P-value is
at most {\it 5\%} for {\it 95\%} of 40,000 simulations,
while running simulations and computing P-values
exactly as for the plots in the present section.
\end{remark}

\begin{remark}
To compute the P-values for each example in Subsection~\ref{with},
we should in principle calculate the maximum-likelihood estimate $\hat\theta$
for each of 40,000 simulations and (for each goodness-of-fit statistic)
use these estimates to perform $(\hbox{40,000})^2$ times
the three-step procedure described in Remark~\ref{MonteCarlo}.
The computational costs for generating the plots
in Subsection~\ref{with} would then be excessive.
Instead, when computing the P-values
as a function of the value of the statistic under consideration,
we calculated $\hat\theta$ only once,
using as the empirical data 1,000,000 draws from the underlying distribution,
and (for each goodness-of-fit statistic) performed 40,000 times
the three-step procedure described in Remark~\ref{MonteCarlo},
using the single value of $\hat\theta$ (but many values of $\tilde\theta$
from Remark~\ref{MonteCarlo}).
The parameter estimates did not vary much over the 40,000 simulations,
so approximating the P-values thus is accurate.
Furthermore, when the parameter is just a permutation,
as in Subsection~\ref{eighthexp},
the ``approximation'' described in the present remark is exactly equivalent
to recomputing the P-values 40,000 times
--- we are not making any approximation at all.
Please note that we did recalculate
the maximum-likelihood estimate $\hat\theta$
(and $\tilde\theta$ from Remark~\ref{MonteCarlo})
for each of 40,000 simulations when computing the values of the statistics
for the simulation; however, when calculating the P-values
as a function of the values of the statistics,
we always drew from the model distribution associated
with the same value of the parameter.
\end{remark}

\begin{remark}
\label{alternatives}
The root-mean-square statistic is not very sensitive to relative discrepancies
between the model and actual distributions
in bins whose associated model probabilities are small.
When sensitivity in these bins is desirable,
we recommend using both the root-mean-square statistic
and an asymptotically equivalent variation of $\chi^2$,
such as the log--likelihood-ratio or ``$G^2$'' statistic.
\end{remark}


\subsection{Examples without parameter estimation}
\label{without}

\subsubsection{A simple, illustrative example}
\label{firstex}

Let us first specify the model distribution to be
\begin{equation}
\label{first}
p_0^{(1)} = \frac{1}{4},
\end{equation}
\begin{equation}
p_0^{(2)} = \frac{1}{4},
\end{equation}
and
\begin{equation}
\label{last}
p_0^{(j)} = \frac{1}{2m-4}
\end{equation}
for $j = 3$,~$4$, \dots,~$m$.
We consider $n$ i.i.d.\ draws from the distribution
\begin{equation}
\label{alt_0}
p^{(1)} = \frac{3}{8},
\end{equation}
\begin{equation}
p^{(2)} = \frac{1}{8},
\end{equation}
and
\begin{equation}
\label{alt_00}
p^{(j)} = p_0^{(j)}
\end{equation}
for $j = 3$,~$4$, \dots,~$m$,
where $p_0^{(3)}$, $p_0^{(4)}$, \dots,~$p_0^{(m)}$ are the same as
in~(\ref{last}).

\begin{figure}
\begin{center}
\rotatebox{-90}{\scalebox{.47}{\includegraphics{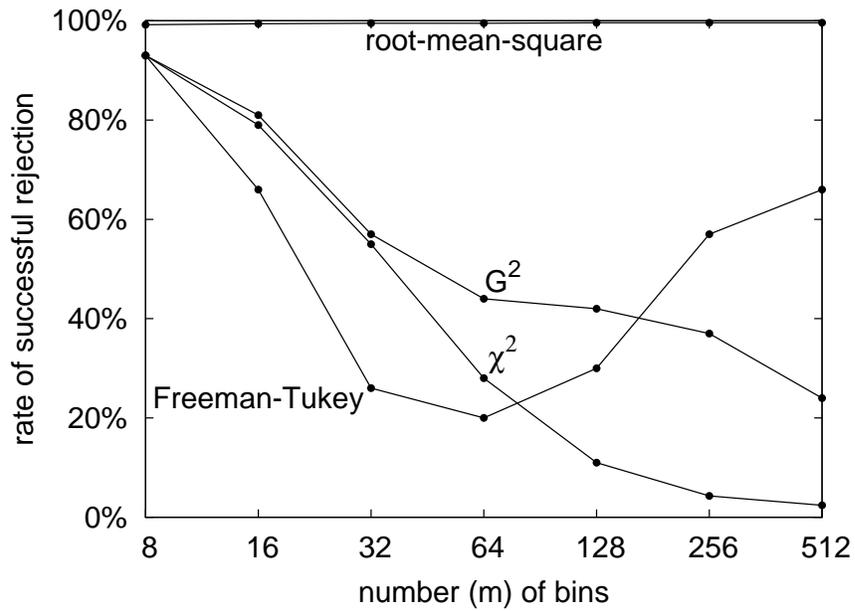}}}
\\\vspace{.1in}
\caption{First example, with $n = 200$ draws; see Subsection~\ref{firstex}.}
\label{plot2}
\end{center}
\end{figure}

Figure~\ref{plot2} plots the percentage of 40,000 simulations,
each generating 200 i.i.d.\ draws according
to the actual distribution defined in~(\ref{alt_0})--(\ref{alt_00}),
that are successfully detected as not arising from the model distribution
at the $1\%$ significance level.
We computed the P-values by conducting 40,000 simulations,
each generating 200 i.i.d.\ draws according
to the model distribution defined in~(\ref{first})--(\ref{last}).
Figure~\ref{plot2} shows that the root-mean-square is successful
in at least 99\% of the simulations,
while the classical $\chi^2$ statistic fails often,
succeeding in less than 80\% of the simulations for $m = 16$,
and less than 5\% for $m \ge 256$.

\begin{figure}
\begin{center}
\rotatebox{-90}{\scalebox{.47}{\includegraphics{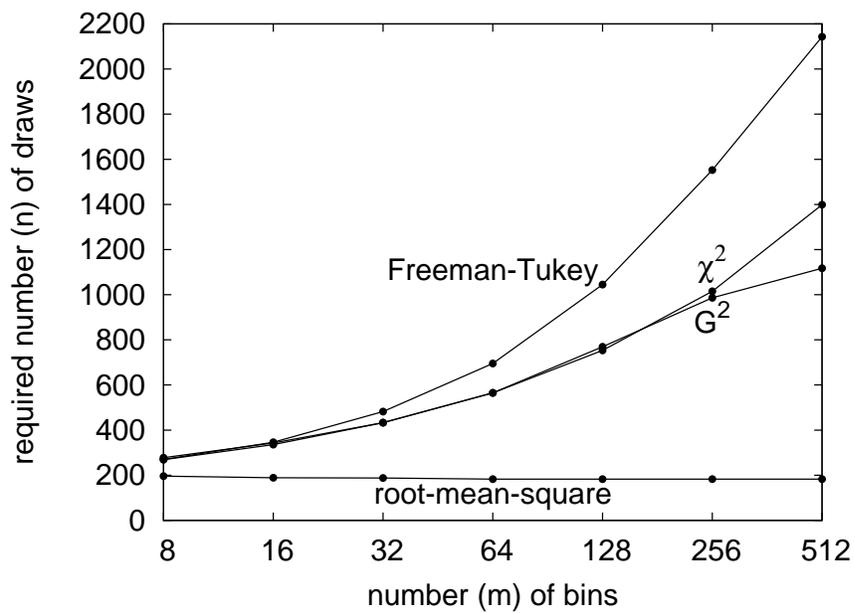}}}
\\\vspace{.1in}
\caption{First example (statistical ``efficiency'');
see Subsection~\ref{firstex}.}
\label{plot}
\end{center}
\end{figure}

Figure~\ref{plot} plots the number $n$ of draws required to distinguish
the actual distribution defined in~(\ref{alt_0})--(\ref{alt_00})
from the model distribution defined in~(\ref{first})--(\ref{last}).
Remark~\ref{distinguish} above specifies what we mean by ``distinguish.''
Figure~\ref{plot} shows that the root-mean-square requires
only about $n = 185$ draws for any number $m$ of bins,
while the classical $\chi^2$ statistic requires
$90\%$ more draws for $m = 16$, and greater than $300\%$ more for $m \ge 128$.
Furthermore, the classical $\chi^2$ statistic requires increasingly many draws
as the number $m$ of bins increases, unlike the root-mean-square.

\newpage

\subsubsection{Truncated power-laws}
\label{secondex}

Next, let us specify the model distribution to be
\begin{equation}
\label{Zipf1}
p_0^{(j)} = \frac{C_1}{j}
\end{equation}
for $j = 1$,~$2$, \dots,~$m$, where
\begin{equation}
\label{const1}
C_1 = \frac{1}{\sum_{j=1}^m 1/j}.
\end{equation}
We consider $n$ i.i.d.\ draws from the distribution
\begin{equation}
\label{Zipf2}
p^{(j)} = \frac{C_2}{j^2}
\end{equation}
for $j = 1$,~$2$, \dots,~$m$, where
\begin{equation}
\label{const2}
C_2 = \frac{1}{\sum_{j=1}^m 1/j^2}.
\end{equation}

\begin{figure}
\begin{center}
\rotatebox{-90}{\scalebox{.47}{\includegraphics{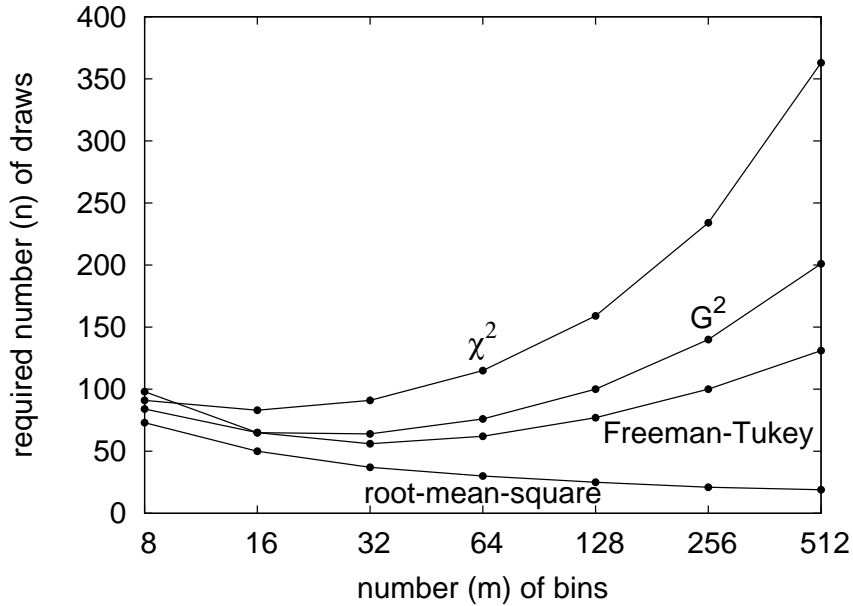}}}
\\\vspace{.1in}
\caption{Second example; see Subsection~\ref{secondex}.}
\label{plotz}
\end{center}
\end{figure}

Figure~\ref{plotz} plots the number $n$ of draws required to distinguish
the actual distribution defined in~(\ref{Zipf2}) and~(\ref{const2})
from the model distribution defined in~(\ref{Zipf1}) and~(\ref{const1}).
Remark~\ref{distinguish} above specifies what we mean by ``distinguish.''
Figure~\ref{plotz} shows that the classical $\chi^2$ statistic
requires increasingly many draws as the number $m$ of bins increases,
while the root-mean-square exhibits the opposite behavior.

\newpage

\subsubsection{Additional truncated power-laws}
\label{thirdex}

Let us again specify the model distribution to be
\begin{equation}
\label{Zipf11}
p_0^{(j)} = \frac{C_1}{j}
\end{equation}
for $j = 1$,~$2$, \dots,~$m$, where
\begin{equation}
\label{const11}
C_1 = \frac{1}{\sum_{j=1}^m 1/j}.
\end{equation}
We now consider $n$ i.i.d.\ draws from the distribution
\begin{equation}
\label{Zipf12}
p^{(j)} = \frac{C_{1/2}}{\sqrt{j}}
\end{equation}
for $j = 1$,~$2$, \dots,~$m$, where
\begin{equation}
\label{const12}
C_{1/2} = \frac{1}{\sum_{j=1}^m 1/\sqrt{j}}.
\end{equation}

\begin{figure}
\begin{center}
\rotatebox{-90}{\scalebox{.47}{\includegraphics{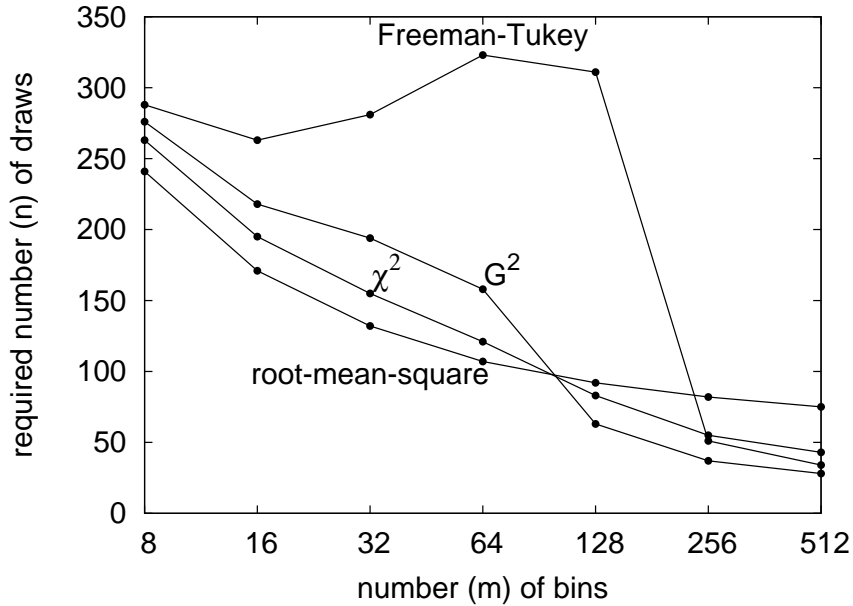}}}
\\\vspace{.1in}
\caption{Third example; see Subsection~\ref{thirdex}.}
\label{plotz2}
\end{center}
\end{figure}

Figure~\ref{plotz2} plots the number $n$ of draws required to distinguish
the actual distribution defined in~(\ref{Zipf12}) and~(\ref{const12})
from the model distribution defined in~(\ref{Zipf11}) and~(\ref{const11}).
Remark~\ref{distinguish} above specifies what we mean by ``distinguish.''
The root-mean-square is not uniformly more powerful than the other statistics
in this example; see Remark~\ref{alternatives}
at the beginning of the present section.

\newpage

\subsubsection{Additional truncated power-laws, reversed}
\label{fourthex}

Let us next specify the model distribution to be
\begin{equation}
\label{Zipf113}
p_0^{(j)} = \frac{C_{1/2}}{\sqrt{j}}
\end{equation}
for $j = 1$,~$2$, \dots,~$m$, where
\begin{equation}
\label{const113}
C_{1/2} = \frac{1}{\sum_{j=1}^m 1/\sqrt{j}}.
\end{equation}
We now consider $n$ i.i.d.\ draws from the distribution
\begin{equation}
\label{Zipf123}
p^{(j)} = \frac{C_1}{j}
\end{equation}
for $j = 1$,~$2$, \dots,~$m$, where
\begin{equation}
\label{const123}
C_1 = \frac{1}{\sum_{j=1}^m 1/j}.
\end{equation}

\begin{figure}
\begin{center}
\rotatebox{-90}{\scalebox{.47}{\includegraphics{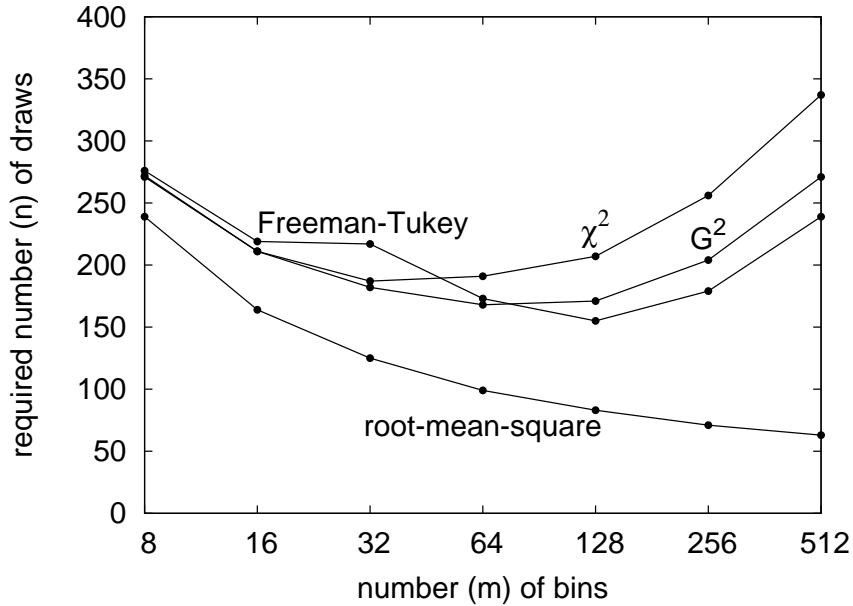}}}
\\\vspace{.1in}
\caption{Fourth example; see Subsection~\ref{fourthex}.}
\label{plotz12}
\end{center}
\end{figure}

Figure~\ref{plotz12} plots the number $n$ of draws required to distinguish
the actual distribution defined in~(\ref{Zipf123}) and~(\ref{const123})
from the model distribution defined in~(\ref{Zipf113}) and~(\ref{const113}).
Remark~\ref{distinguish} above specifies what we mean by ``distinguish.''
Figure~\ref{plotz12} shows that the classical $\chi^2$ statistic
requires many times more draws than the root-mean-square,
as the number $m$ of bins increases.

\newpage

\subsubsection{A final example with fully specified truncated power-laws}
\label{fifthex}

Let us next specify the model distribution to be
\begin{equation}
\label{Zipf115}
p_0^{(j)} = \frac{C_2}{j^2}
\end{equation}
for $j = 1$,~$2$, \dots,~$m$, where
\begin{equation}
\label{const115}
C_2 = \frac{1}{\sum_{j=1}^m 1/j^2}.
\end{equation}
We again consider $n$ i.i.d.\ draws from the distribution
\begin{equation}
\label{Zipf125}
p^{(j)} = \frac{C_1}{j}
\end{equation}
for $j = 1$,~$2$, \dots,~$m$, where
\begin{equation}
\label{const125}
C_1 = \frac{1}{\sum_{j=1}^m 1/j}.
\end{equation}

\begin{figure}
\begin{center}
\rotatebox{-90}{\scalebox{.47}{\includegraphics{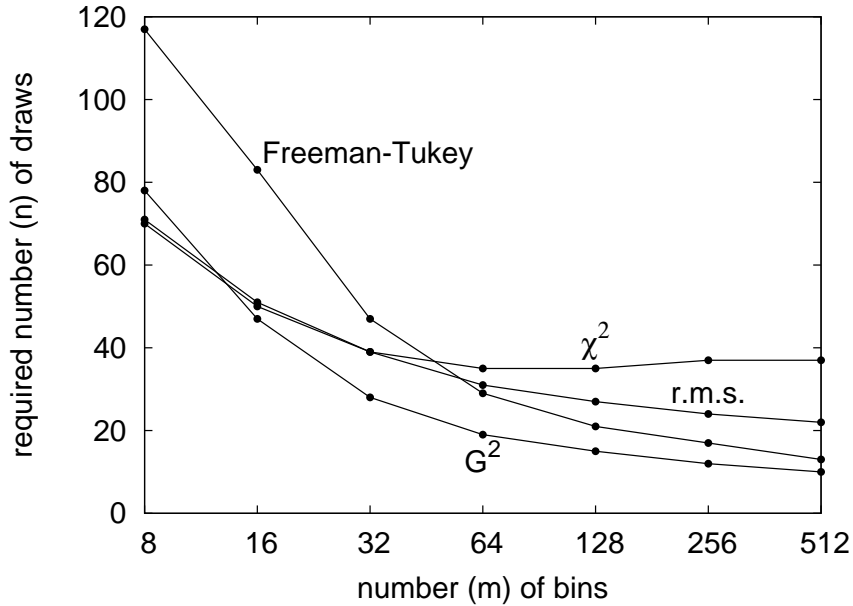}}}
\\\vspace{.1in}
\caption{Fifth example; see Subsection~\ref{fifthex}.}
\label{plotzz}
\end{center}
\end{figure}

Figure~\ref{plotzz} plots the number $n$ of draws required to distinguish
the actual distribution defined in~(\ref{Zipf125}) and~(\ref{const125})
from the model distribution defined in~(\ref{Zipf115}) and~(\ref{const115}).
Remark~\ref{distinguish} above specifies what we mean by ``distinguish.''
The root-mean-square is not uniformly more powerful than the other statistics
in this example; see Remark~\ref{alternatives}
at the beginning of the present section.

\newpage

\subsubsection{Modified Poisson distributions}
\label{sixthex}

Let us specify the model distribution to be
the (truncated) Poisson distribution
\begin{equation}
\label{wpoisson1alt}
p_0^{(j)} = \frac{B_{3m/8} \, \left(\frac{3m}{8}\right)^{j-1}}{(j-1)!}
\end{equation}
for $j = 1$,~$2$, \dots,~$m$, where
\begin{equation}
\label{wpoisson2alt}
B_{3m/8} = \frac{1}{\sum_{j=1}^m \left(\frac{3m}{8}\right)^{j-1}/(j-1)!}.
\end{equation}
We consider $n$ i.i.d.\ draws from the distribution
\begin{equation}
\label{wpoisson1a}
p^{((3m/8)-1)} = S/10,
\end{equation}
\begin{equation}
\label{wpoisson2a}
p^{(3m/8)} = 4S/5,
\end{equation}
\begin{equation}
\label{wpoisson3a}
p^{((3m/8)+1)} = S/10,
\end{equation}
\begin{equation}
\label{wpoisson4a}
S = p_0^{((3m/8)-1)} + p_0^{(3m/8)} + p_0^{((3m/8)+1)},
\end{equation}
\begin{equation}
\label{wpoisson5a}
p^{(j)} = p_0^{(j)}
\end{equation}
for the remaining values of $j$
(for $j = 1$,~$2$, \dots,~$\frac{3m}{8}-2$
and $j = \frac{3m}{8}+2$,~$\frac{3m}{8}+3$, \dots,~$m$).

\begin{figure}
\begin{center}
\rotatebox{-90}{\scalebox{.47}{\includegraphics{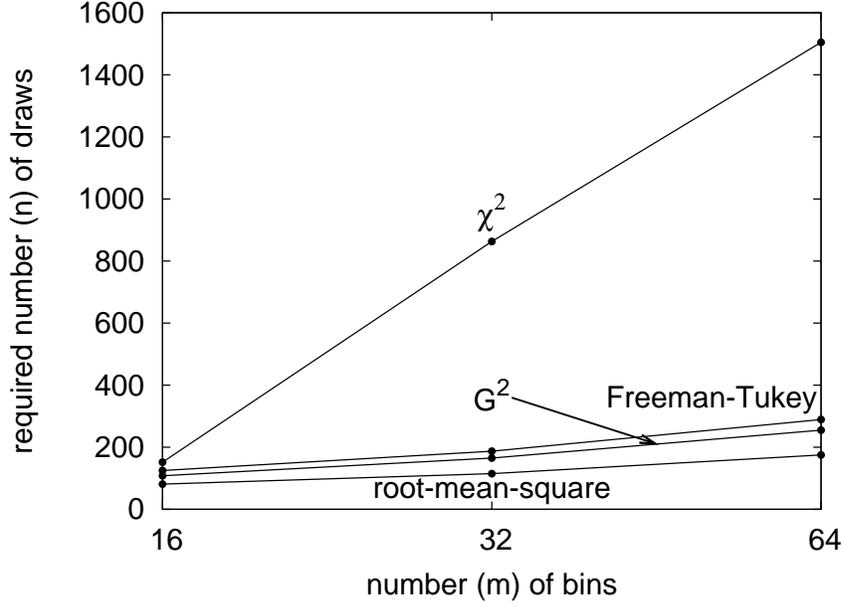}}}
%
%
%
\caption{Sixth example; see Subsection~\ref{sixthex}.}
\label{plotp}
\end{center}
\vspace{-.065in}
\end{figure}

Figure~\ref{plotp} plots the number $n$ of draws required to distinguish
the actual distribution defined in~(\ref{wpoisson1a})--(\ref{wpoisson5a})
from the model distribution defined
in~(\ref{wpoisson1alt}) and~(\ref{wpoisson2alt}).
Remark~\ref{distinguish} above specifies what we mean by ``distinguish.''

\newpage

\subsubsection{A truncated power-law and a truncated geometric distribution}
\label{seventhex}

\begin{figure}
\begin{center}
\rotatebox{-90}{\scalebox{.47}{\includegraphics{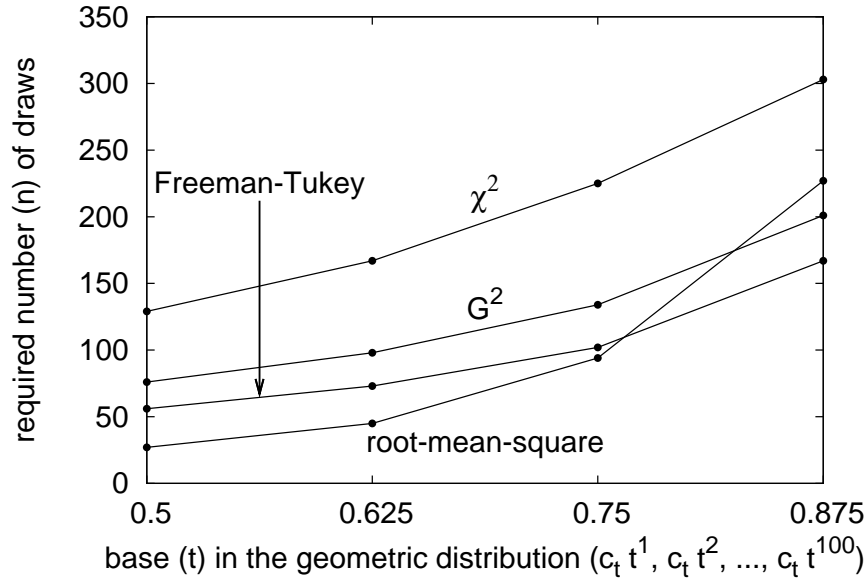}}}
\\\vspace{.1in}
\caption{Seventh example; see Subsection~\ref{seventhex}.}
\label{plotg}
\end{center}
%
%
\end{figure}

Let us finally specify the model distribution to be
\begin{equation}
\label{wZipf}
p_0^{(j)} = \frac{C_1}{j}
\end{equation}
for $j = 1$,~$2$, \dots,~$100$, where
\begin{equation}
\label{wconst}
C_1 = \frac{1}{\sum_{j=1}^{100} 1/j}.
\end{equation}
We consider $n$ i.i.d.\ draws from the (truncated) geometric distribution
\begin{equation}
\label{wgeom}
p^{(j)} = c_t \, t^j
\end{equation}
for $j = 1$,~$2$, \dots,~$100$, where
\begin{equation}
\label{wconstt}
c_t = \frac{1}{\sum_{j=1}^{100} t^j};
\end{equation}
Figure~\ref{plotg} considers several values for $t$.

Figure~\ref{plotg} plots the number $n$ of draws required to distinguish
the actual distribution defined in~(\ref{wgeom}) and~(\ref{wconstt})
from the model distribution defined in~(\ref{wZipf}) and~(\ref{wconst}).
Remark~\ref{distinguish} above specifies what we mean by ``distinguish.''
See the next section, Subsection~\ref{firstexp}, for a similar example,
this time involving parameter estimation.

\newpage

\begin{figure}
\begin{center}
\rotatebox{-90}{\scalebox{.47}{\includegraphics{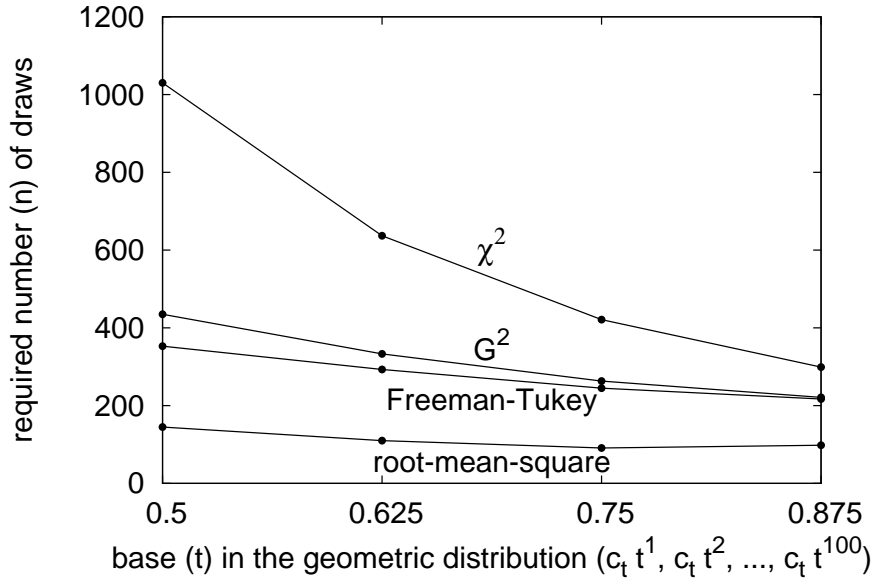}}}
\\\vspace{.1in}
\caption{First example; see Subsection~\ref{firstexp}.}
\label{plotparz}
\end{center}
\end{figure}

\subsection{Examples with parameter estimation}
\label{with}

\subsubsection{A truncated power-law and a truncated geometric distribution}
\label{firstexp}

We turn now to models involving parameter estimation,
as detailed by~\cite{perkins-tygert-ward2}.
Let us specify the model distribution to be the Zipf distribution
\begin{equation}
\label{Zipftheta}
p_0^{(j)}(\theta) = \frac{C_{\theta}}{j^{\theta}}
\end{equation}
for $j = 1$,~$2$, \dots,~$100$, where
\begin{equation}
\label{consttheta}
C_{\theta} = \frac{1}{\sum_{j=1}^{100} 1/j^{\theta}};
\end{equation}
we estimate the parameter $\theta$ via maximum-likelihood methods.
We consider $n$ i.i.d.\ draws from the (truncated) geometric distribution
\begin{equation}
\label{geom}
p^{(j)} = c_t \, t^j
\end{equation}
for $j = 1$,~$2$, \dots,~$100$, where
\begin{equation}
\label{constt}
c_t = \frac{1}{\sum_{j=1}^{100} t^j};
\end{equation}
Figure~\ref{plotparz} considers several values for $t$.

Figure~\ref{plotparz} plots the number $n$ of draws required to distinguish
the actual distribution defined in~(\ref{geom}) and~(\ref{constt})
from the model distribution defined in~(\ref{Zipftheta}) and~(\ref{consttheta}),
estimating the parameter $\theta$ in~(\ref{Zipftheta}) and~(\ref{consttheta})
via maximum-likelihood methods.
Remark~\ref{distinguish} above specifies what we mean by ``distinguish.''

\newpage

\begin{figure}
\begin{center}
\rotatebox{-90}{\scalebox{.47}{\includegraphics{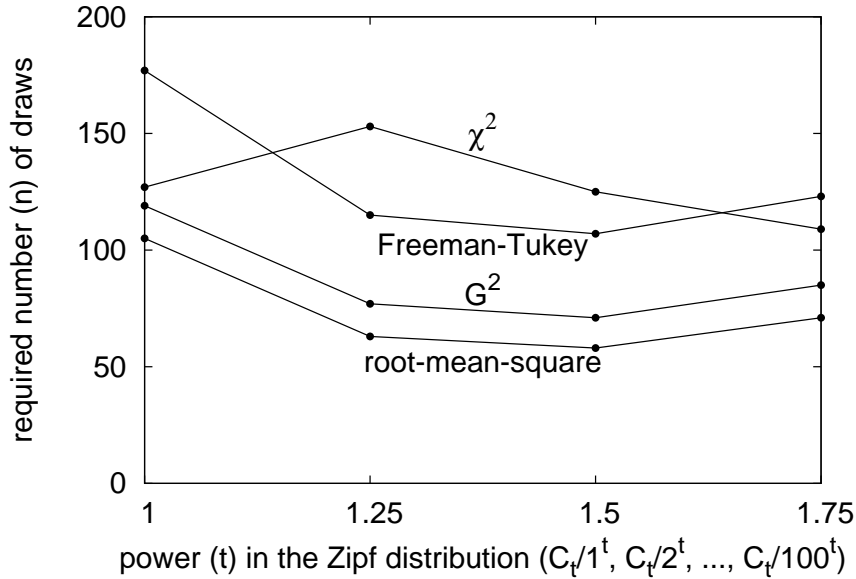}}}
\\\vspace{.1in}
\caption{Second example; see Subsection~\ref{secondexp}.}
\label{plotparg}
\end{center}
\end{figure}

\subsubsection{A rebinned geometric distribution and a truncated power-law}
\label{secondexp}

Let us specify the model distribution to be
\begin{equation}
\label{geomalt1}
p_0^{(j)}(\theta) = \theta^{j-1} (1-\theta)
\end{equation}
for $j = 1$,~$2$, \dots,~$99$, and
\begin{equation}
\label{geomalt2}
p_0^{(100)}(\theta) = \theta^{99};
\end{equation}
we estimate the parameter $\theta$ via maximum-likelihood methods.
We consider $n$ i.i.d.\ draws from the Zipf distribution
\begin{equation}
\label{Zipft}
p^{(j)} = \frac{C_t}{j^t}
\end{equation}
for $j = 1$,~$2$, \dots,~$100$, where
\begin{equation}
\label{const}
C_t = \frac{1}{\sum_{j=1}^{100} 1/j^t};
\end{equation}
Figure~\ref{plotparg} considers several values for $t$.

Figure~\ref{plotparg} plots the number $n$ of draws required to distinguish
the actual distribution defined in~(\ref{Zipft}) and~(\ref{const})
from the model distribution defined in~(\ref{geomalt1}) and~(\ref{geomalt2}),
estimating the parameter $\theta$ in~(\ref{geomalt1}) and~(\ref{geomalt2})
via maximum-likelihood methods.
Remark~\ref{distinguish} above specifies what we mean by ``distinguish.''

\newpage

\begin{figure}
\begin{center}
\rotatebox{-90}{\scalebox{.47}{\includegraphics{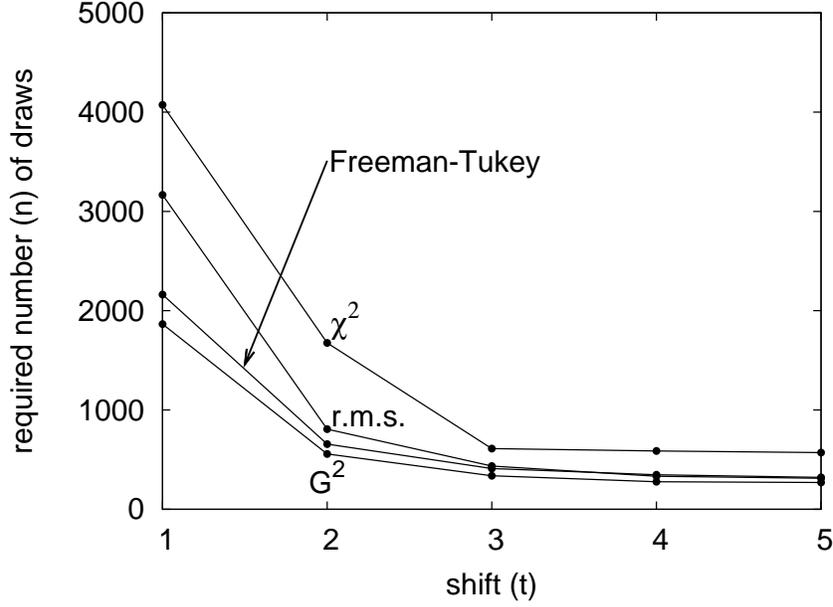}}}
\\\vspace{.1in}
\caption{Third example; see Subsection~\ref{thirdexp}.}
\label{plotparps}
\end{center}
\end{figure}

\subsubsection{Truncated shifted Poisson distributions}
\label{thirdexp}

Let us specify the model distribution to be
the (truncated) Poisson distribution
\begin{equation}
\label{poisson1}
p_0^{(j)}(\theta) = \frac{B_{\theta} \, \theta^{j-1}}{(j-1)!}
\end{equation}
for $j = 1$,~$2$, \dots,~$21$, where
\begin{equation}
\label{poisson2}
B_{\theta} = \frac{1}{\sum_{j=1}^{21} \theta^{j-1}/(j-1)!};
\end{equation}
we estimate the parameter $\theta$ via maximum-likelihood methods.
We consider $n$ i.i.d.\ draws from the distribution
\begin{equation}
\label{poisson1s}
p^{(j)} = \frac{\tilde{B}_t \, 5^{j-1+t}}{(j-1+t)!}
\end{equation}
for $j = 1$,~$2$, \dots,~$21$, where
\begin{equation}
\label{poisson2s}
\tilde{B}_t = \frac{1}{\sum_{j=1}^{21} 5^{j-1+t}/(j-1+t)!};
\end{equation}
Figure~\ref{plotparps} considers several values for $t$.
Clearly, $p^{(j)} = p_0^{(j)}(5)$ for $j = 1$,~$2$, \dots,~$21$,
if $t = 0$.

Figure~\ref{plotparps} plots the number $n$ of draws required to distinguish
the actual distribution defined in~(\ref{poisson1s}) and~(\ref{poisson2s})
from the model distribution defined in~(\ref{poisson1}) and~(\ref{poisson2}),
estimating the parameter $\theta$ in~(\ref{poisson1}) and~(\ref{poisson2})
via maximum-likelihood methods.
Remark~\ref{distinguish} above specifies what we mean by ``distinguish.''

\newpage

\begin{figure}
\begin{center}
\rotatebox{-90}{\scalebox{.47}{\includegraphics{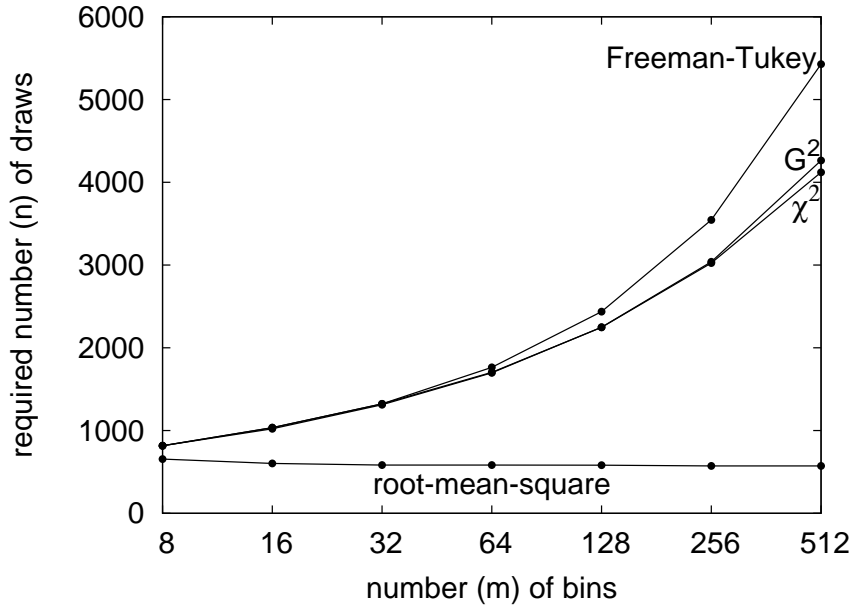}}}
%
%
%
\caption{Fourth example; see Subsection~\ref{sixthexp}.}
\label{plotcc}
\end{center}
\vspace{-.1in}
\end{figure}

\subsubsection{An example with a uniform tail}
\label{sixthexp}

Let us specify the model distribution to be
\begin{equation}
\label{castle1}
p_0^{(1)}(\theta) = \theta,
\end{equation}
\begin{equation}
\label{castle2}
p_0^{(2)}(\theta) = \theta,
\end{equation}
\begin{equation}
\label{castle3}
p_0^{(3)}(\theta) = \frac{1}{2} - 2\theta,
\end{equation}
\begin{equation}
\label{castle4}
p_0^{(j)}(\theta) = \frac{1}{2m-6}
\end{equation}
for $j = 4$,~$5$, \dots,~$m$;
we estimate the parameter $\theta$ via maximum-likelihood methods.
We consider $n$ i.i.d.\ draws from the distribution
\begin{equation}
\label{castle1a}
p^{(1)} = \frac{1}{4},
\end{equation}
\begin{equation}
\label{castle2a}
p^{(2)} = \frac{1}{8},
\end{equation}
\begin{equation}
\label{castle3a}
p^{(3)} = \frac{1}{8},
\end{equation}
\begin{equation}
\label{castle4a}
p^{(j)} = \frac{1}{2m-6}
\end{equation}
for $j = 4$,~$5$, \dots,~$m$.

Figure~\ref{plotcc} plots the number $n$ of draws required to distinguish
the actual distribution defined in~(\ref{castle1a})--(\ref{castle4a})
from the model distribution defined in~(\ref{castle1})--(\ref{castle4}),
estimating the parameter $\theta$ in~(\ref{castle1})--(\ref{castle4})
via maximum-likelihood methods.
Remark~\ref{distinguish} above specifies what we mean by ``distinguish.''

\newpage

\begin{figure}
\begin{center}
\rotatebox{-90}{\scalebox{.47}{\includegraphics{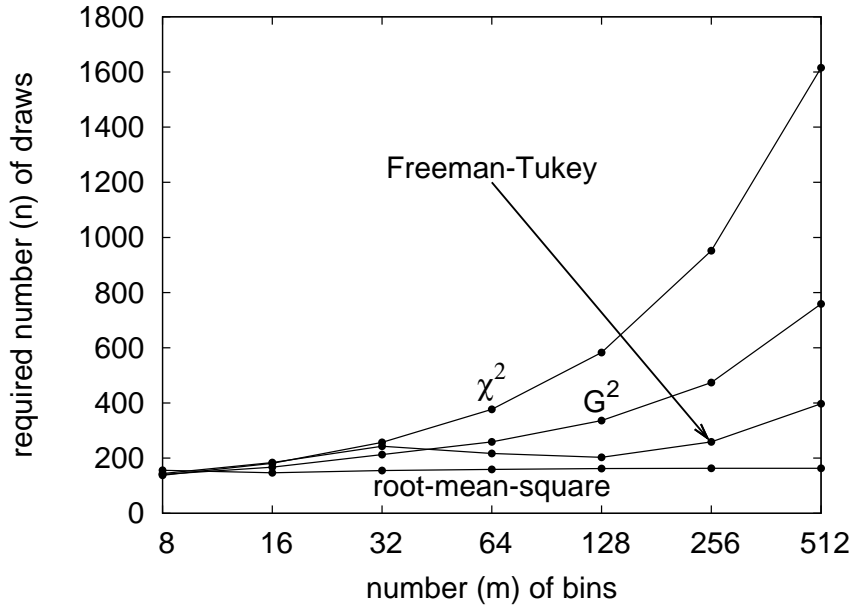}}}
\\\vspace{.1in}
\caption{Fifth example; see Subsection~\ref{seventhexp}.}
\label{plotcu}
\end{center}
\end{figure}

\subsubsection{A model with an integer-valued parameter}
\label{seventhexp}

Let us specify the model distribution to be
\begin{equation}
\label{cu1}
p_0^{(j)}(\theta) = \frac{1}{2\theta}
\end{equation}
for $j = 1$,~$2$, \dots,~$\theta$, and
\begin{equation}
\label{cu2}
p_0^{(j)}(\theta) = \frac{1}{2(m-\theta)}
\end{equation}
for $j = \theta+1$,~$\theta+2$, \dots,~$m$;
we estimate the parameter $\theta$ via maximum-likelihood methods.
We consider $n$ i.i.d.\ draws from the distribution
\begin{equation}
\label{cu1a}
p^{(1)} = \frac{1}{4},
\end{equation}
\begin{equation}
\label{cu2a}
p^{(2)} = \frac{1}{4},
\end{equation}
\begin{equation}
\label{cu3a}
p^{(3)} = \frac{1}{4},
\end{equation}
and
\begin{equation}
\label{cu4a}
p^{(j)} = \frac{1}{4m-12}
\end{equation}
for $j = 4$,~$5$, \dots,~$m$.

Figure~\ref{plotcu} plots the number $n$ of draws required to distinguish
the actual distribution defined in~(\ref{cu1a})--(\ref{cu4a})
from the model distribution defined in~(\ref{cu1}) and~(\ref{cu2}),
estimating the parameter $\theta$ in~(\ref{cu1}) and~(\ref{cu2})
via maximum-likelihood methods.
Remark~\ref{distinguish} above specifies what we mean by ``distinguish.''

\newpage

\begin{figure}
\begin{center}
\rotatebox{-90}{\scalebox{.47}{\includegraphics{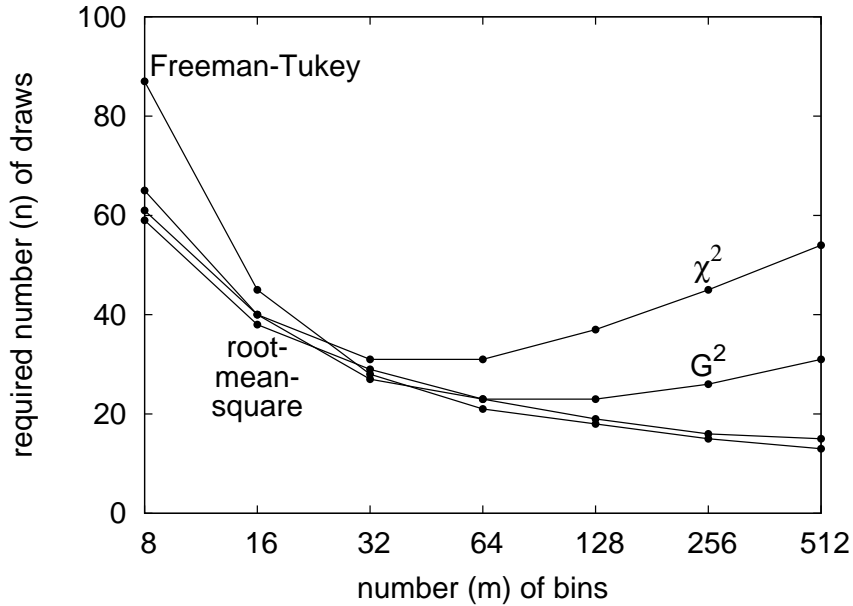}}}
%
%
%
%
\caption{Sixth example; see Subsection~\ref{eighthexp}.}
\label{plotperm}
\vspace{-.125in}
\end{center}
\end{figure}

\subsubsection{Truncated power-laws parameterized with a permutation}
\label{eighthexp}

Let us specify the model to be the Zipf distribution
\begin{equation}
\label{zipfperm}
p_0^{(j)}(\theta) = \frac{C_1}{\theta(j)}
\end{equation}
for $j = 1$,~$2$, \dots,~$m$, where
$\theta$ is a permutation of the integers $1$,~$2$, \dots,~$m$, and
\begin{equation}
\label{zipfpermc}
C_1 = \frac{1}{\sum_{j=1}^m 1/j};
\end{equation}
we estimate the permutation $\theta$ via maximum-likelihood methods, that is,
by sorting the frequencies:
first we choose $j_1$ to be the number of a bin containing
the greatest number of draws among all $m$ bins,
then we choose $j_2$ to be the number of a bin containing
the greatest number of draws among the remaining $m-1$ bins,
then we choose $j_3$ to be the number of a bin containing
the greatest among the remaining $m-2$ bins, and so on,
and finally we find $\theta$ such that $\theta(j_1) = 1$,~$\theta(j_2) = 2$,
\dots,~$\theta(j_m) = m$.

We consider $n$ i.i.d.\ draws from the distribution
\begin{equation}
\label{perma2}
p^{(j)} = \frac{C_2}{j^2}
\end{equation}
for $j = 1$,~$2$, \dots,~$m$, where
\begin{equation}
\label{perma2c}
C_2 = \frac{1}{\sum_{j=1}^m 1/j^2}.
\end{equation}

Figure~\ref{plotperm} plots the number $n$ of draws required to distinguish
the actual distribution defined in~(\ref{perma2}) and~(\ref{perma2c})
from the model distribution defined in~(\ref{zipfperm}) and~(\ref{zipfpermc}),
estimating the parameter $\theta$ in~(\ref{zipfperm})
via maximum-likelihood methods (that is, by sorting).
Remark~\ref{distinguish} above specifies what we mean by ``distinguish.''

\newpage

\begin{figure}
\begin{center}
\rotatebox{-90}{\scalebox{.47}{\includegraphics{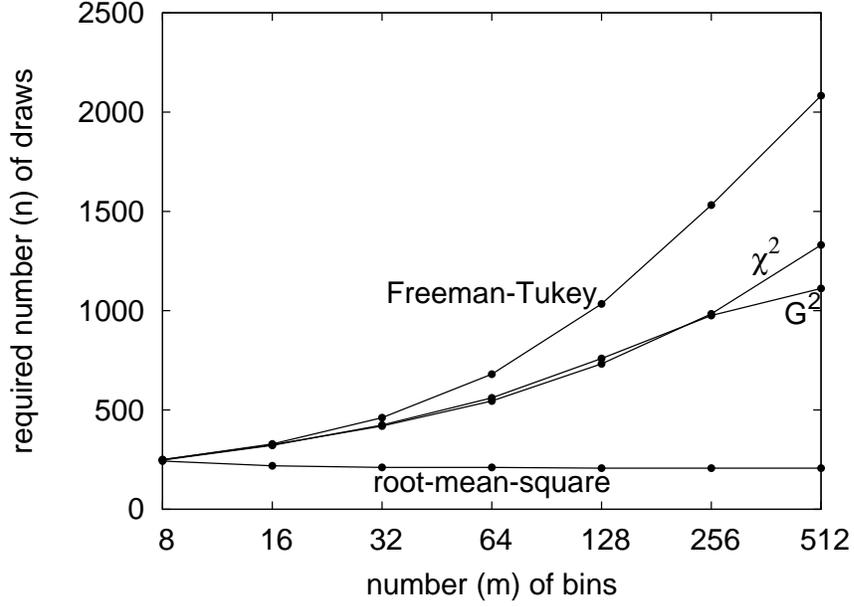}}}
\\\vspace{.1in}
\caption{Seventh example; see Subsection~\ref{tenthexp}.}
\label{plotpar2}
\end{center}
\end{figure}

\subsubsection{A model with two parameters}
\label{tenthexp}

For the final example, let us specify the model distribution to be
\begin{equation}
\label{two1}
p_0^{(1)}(\theta_1,\theta_2) = \theta_1,
\end{equation}
\begin{equation}
\label{two2}
p_0^{(2)}(\theta_1,\theta_2) = \theta_1,
\end{equation}
\begin{equation}
\label{two3}
p_0^{(3)}(\theta_1,\theta_2) = \theta_2,
\end{equation}
\begin{equation}
\label{two4}
p_0^{(4)}(\theta_1,\theta_2) = \theta_2,
\end{equation}
and
\begin{equation}
\label{two5}
p_0^{(j)}(\theta_1,\theta_2) = \frac{1-2\theta_1-2\theta_2}{m-4}
\end{equation}
for $j = 5$,~$6$, \dots,~$m$;
we estimate the parameters $\theta_1$ and~$\theta_2$
via maximum-likelihood methods.
We consider $n$ i.i.d.\ draws from the distribution
\begin{equation}
\label{two1a}
p^{(1)} = \frac{9}{32},
\end{equation}
\begin{equation}
\label{two2a}
p^{(2)} = \frac{3}{32},
\end{equation}
\begin{equation}
\label{two3a}
p^{(3)} = \frac{3}{32},
\end{equation}
\begin{equation}
\label{two4a}
p^{(4)} = \frac{1}{32},
\end{equation}
and
\begin{equation}
\label{two5a}
p^{(j)} = \frac{1}{2m-8}
\end{equation}
for $j = 5$,~$6$, \dots,~$m$.

Figure~\ref{plotpar2} plots the number $n$ of draws required to distinguish
the actual distribution defined in~(\ref{two1a})--(\ref{two5a})
from the model distribution defined in~(\ref{two1})--(\ref{two5}),
estimating the parameters $\theta_1$ and~$\theta_2$
in~(\ref{two1})--(\ref{two5}) via maximum-likelihood methods.
Remark~\ref{distinguish} above specifies what we mean by ``distinguish.''

\section*{Acknowledgements}

We would like to thank Alex Barnett, G\'erard Ben Arous, James Berger,
Tony Cai, Sourav Chatterjee, Ronald Raphael Coifman, Ingrid Daubechies,
Jianqing Fan, Andrew Gelman, Leslie Greengard, Peter W. Jones, Deborah Mayo,
Peter McCullagh, Michael O'Neil, Ron Peled, William Press, Vladimir Rokhlin,
Joseph Romano, Gary Simon, Amit Singer, Michael Stein, Stephen Stigler,
Joel Tropp, Larry Wasserman, and Douglas A. Wolfe.
We would also like to thank Jiayang Gao for her many observations,
which include pointing out the identity in~(\ref{freeman-tukey}),
showing that the Freeman-Tukey/Hellinger-distance statistic
is just a weighted version of the root-mean-square.

William Perkins was supported in part by NSF Grant OISE-0730136
and an NSF Postdoctoral Research Fellowship.
Mark Tygert was supported in part by a Research Fellowship
from the Alfred P. Sloan Foundation.
Rachel Ward was supported in part by an NSF Postdoctoral Research Fellowship
and a Donald D. Harrington Faculty Fellowship.


\setlength{\bibsep}{4.6pt}
\bibliographystyle{asamod.bst}
\bibliography{stat}

\end{document}